# Experimental investigation of the residues produced in the $^{136}$Xe+Pb and $^{124}$Xe+Pb fragmentation reactions at 1 $A$ GeV

D. Henzlova[1,*,†,‡], K.-H. Schmidt[1], M.V. Ricciardi[1], A. Kelić[1], V. Henzl[1,*], P. Napolitani[1,2], L. Audouin[3], J. Benlliure[4], A. Boudard[5], E. Casarejos[4], J. E. Ducret[5], T. Enqvist[6], A. Heinz[7], A. Junghans[8], B. Jurado[9], A. Krása[10], T. Kurtukian[4], S. Leray[5], M. F. Ordóñez[4,§], J. Pereira[4,*], R. Pleskač[1], F. Rejmund[11], C. Schmitt[12], C. Stéphan[3], L. Tassan-Got[3], C. Villagrasa[5], C. Volant[5], A. Wagner[10], and O. Yordanov[1,**]

[1] GSI, Planckstraße 1, 64291 Darmstadt, Germany
[2] LPC Caen, Université de Caen, CNRS/IN2P3, ENSICAEN, Caen, France
[3] IPN Orsay, IN2P3, 91406 Orsay, France
[4] University of Santiago de Compostela, 15706 Santiago de Compostela, Spain
[5] DAPNIA/SPhN CEA/Saclay, 91191 Gif-sur-Yvette, France
[6] CUPP Project, P. O. Box 22, 86801, Pyhäsalmi, Finland
[7] Wright Nuclear Structure Laboratory, Yale University, New Heaven, CT 06520, USA
[8] Forschungszentrum Rossendorf, 01314 Dresden, Germany
[9] CENBG, IN2P3, 33175 Gradignan, France
[10] Nuclear Physics Institute, 25068 Řež, Czech Republic
[11] GANIL, 14076 Caen, France
[12] Université Lyon I, CNRS/IN2P3, IPNL, Rue Enrico Fermi, 69622 Villeurbanne, France

**Abstract**
The nuclide cross sections and the longitudinal velocity distributions of residues produced in the reactions of $^{136}$Xe and $^{124}$Xe at 1 $A$ GeV in a lead target were measured at the high-resolution magnetic spectrometer, the Fragment Separator (FRS) of GSI. The data cover a broad range of isotopes of the elements between $Z = 3$ and $Z = 56$ for $^{136}$Xe and between $Z = 5$ and $Z = 55$ for $^{124}$Xe, reaching down to cross sections of a few microbarns. The velocity distributions exhibit a Gaussian shape for masses above $A = 20$, while more complex behaviour is observed for lighter masses. The isotopic distributions for both reactions preserve a memory on the projectile $N/Z$ ratio over the whole residue mass range.

# I Introduction

Heavy-ion collisions, being an ideal tool for producing hot nuclear matter at various densities in the laboratory, are an important source of information on the properties of nuclear matter under extreme conditions.

Experimental studies in this field extend from the Fermi-energy regime to relativistic energies. These two energy regimes are characterised by quite different types of reaction dynamics: If the projectile velocity is comparable with the Fermi velocity, nucleon exchange between the reaction partners during the collision is important. During the collision stage, the neutron-to-proton ratio $N/Z$ of projectile and target tends to equilibrate. Observation of the magnitude of this process (known as isospin diffusion) carries information on the symmetry energy coefficient at high temperature and density [1,2]. Non-central collisions at these energies are characterised by deep-inelastic transfer [3], while central collisions form a highly excited compressed piece of nuclear matter, suitable for the study of multifragment decay and phase transitions [4].

---

[*] Present address: NSCL-MSU, 1 Cyclotron, East Lansing, MI 48824, USA
[†] This work forms part of the PhD Thesis of Daniela Henzlova
[‡] Corresponding author: henzlova@nscl.msu.edu
[§] Present address: CIEMAT, Avda. Complutense 22, 28040 Madrid, Spain
[**] Present address: INRNE, 72 Tzarigradsko chaussee, BG-1784 Sofia, Bulgaria

At relativistic energies, the Fermi spheres of projectile and target are well separated. Transfer of nucleons is very unlikely, and the geometrical abrasion picture seems well justified [5]. Flow patterns of nucleons and particles as well as kaon production from central collisions have been studied to deduce the equation of state of hot and compressed nuclear matter [6,7]. At larger impact parameters, spectator matter is sheared off from projectile and target and continues moving essentially with its original velocity. A considerable amount of excitation energy [8] and a slight momentum transfer [9] are induced, but compression is small. Thus, peripheral heavy-ion collisions at relativistic energies are ideal scenario for studying multifragment decay due to purely thermal instabilities [10], avoiding compression effects.

Most experimental devices, developed for such studies, aim for covering the full solid angle in order to provide a complete survey on multiplicities and correlations of charged-particle production. Although this information is crucial for many conclusions, high-precision information on mass and momentum is lacking, since the resolution of these devices is rather limited. This is particularly unsatisfactory if the evolution of the $N/Z$ degree of freedom is to be studied. In order to study effects of "isospin diffusion", systematic studies on the isotopic distributions of heavy nuclei from heavy-ion collisions at Fermi energies have been performed with high-resolution spectrometers, e.g. [11], using projectiles and targets with different neutron excess. However, the situation after the collision stage is not directly observable, because it is further modified by sequential and eventually multifragment decay. Consequently, the final observed isotopic distributions carry the combined information from both – collision and decay stage of the reaction. On the other hand, the isotopic distributions of heavy nuclei from relativistic heavy-ion collisions, where nucleon exchange during the collision is negligible, can give more direct insight into the influence of sequential decay and eventually multifragment decay on the $N/Z$ degree of freedom. Therefore, combining the experimental results from the two energy regimes one can gain more specific information on the evolution of the $N/Z$ degree of freedom during the different stages of the reaction.

Moreover, high-resolution experiments on heavy-ion collisions at relativistic energies can provide deeper insight into the thermal properties of highly excited nuclear systems and on possible liquid-gas phase transitions [12,13]. Detailed study of the kinematics of the spectator matter has also revealed information on the non-local properties of the nuclear force [14,15].

Unfortunately, detailed experimental information on heavy-nuclide production at relativistic energies is rather lacking. An experimental campaign on the formation of heavy residues in proton- and deuteron-induced spallation reactions performed at GSI, which was motivated by nuclear technology, has demonstrated the power of the experimental method and has provided a systematic set of high-quality data [16]. Similar data on heavy-ion collisions are much scarcer, although using the heavy-ion accelerator at GSI (SIS18) considerably higher excitation energies can be introduced into the colliding system. Thus, our knowledge on the properties of highly excited nuclei may be appreciably extended.

The present work is dedicated to the first experiment on determining the full nuclide distributions of projectile-like residues in two systems with very different $N/Z$ content: $^{124}$Xe and $^{136}$Xe + Pb, at the beam energy of 1 $A$ GeV. The heavy target nucleus has been chosen in order to minimize the surface that divides the projectile spectator from the participant zone for a given size of the projectile spectator. Thus, surface effects due to the abrasion process should be minimised. The experiment has been performed at the magnetic spectrometer FRS at GSI. Due to its high resolution, all residues were identified in mass number $A$ and nuclear charge $Z$. Moreover, the longitudinal momenta of all residues were determined with high precision.

The present paper explores the isotopic composition of the final residues measured in a broad range of nuclear charge and the influence of the neutron-to-proton ratio of the initial system on the isotopic composition of the final residues. Results on high-resolution measurements of longitudinal residue velocities are also given. This paper is dedicated to present the experimental data. More detailed discussions and comparisons with nuclear-reaction models will be the subjects of forthcoming publications.

## II Experimental technique

The experiments with the two xenon beams, $^{124}$Xe and $^{136}$Xe, both at 1 $A$ GeV, were performed at the Fragment Separator (FRS) at GSI. The beams were delivered by the heavy-ion synchrotron (SIS) with intensities of ~3·10$^8$ and ~4·10$^7$ particles per spill for $^{124}$Xe and $^{136}$Xe, respectively. In both experiments, the spill length was varied between 1 and 10 sec in order to keep the maximum counting rate allowed by the detector limits and the data acquisition system. In the following, the principles of the isotopic identification at the Fragment Separator and its detection system will be described.

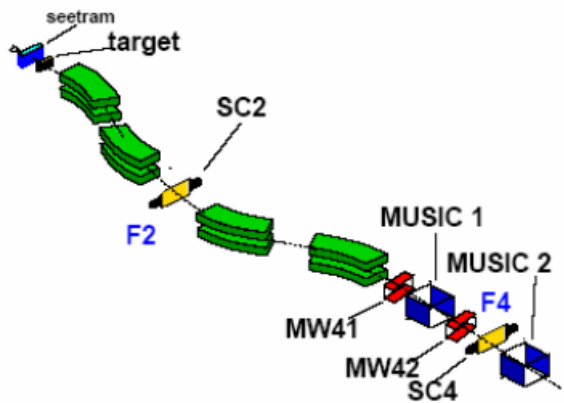

**Figure 1:** (Color online) Schematic view of the Fragment Separator (FRS) and the associated detector equipment. See text for details.

The Fragment Separator [17] is a high-resolution magnetic spectrometer, which allows mass and element separation of the final residues ranging from the lightest masses up to the mass of the heavy projectile. The projectile-like fragments exiting the target are detected and isotopically identified in flight in the associated detector equipment. A schematic view of the experimental setup is shown in Fig. 1.

The FRS is a two-stage magnetic spectrometer with a maximum bending power of 18 Tm, an angular acceptance of 15 mrad around the beam axis, and a momentum acceptance of 3%. A beam monitor, SEETRAM (Secondary Electron TRAnsmission Monitor), was used to constantly measure the number of incoming beam particles. The beam monitor and its calibration system are mounted in front of the spectrometer [18,19,20]. The SEETRAM consists of three thin foils of 11.5 cm in diameter which are mounted perpendicular to the beam axis. The outer foils are made of 14 μm thin aluminium, while the middle foil is made of a 10 μm thin titanium layer. The reaction rate in the SEETRAM amounts to less than 0.1% and the energy loss to less than 0.05% for both primary beams. The target is located 2.27 m in front of the first quadrupole of the FRS. In both experiments a natural lead foil of 635 mg/cm$^2$ thickness was used as a target. The primary beam looses less than 2% of its energy in the lead target. Corrections due to energy loss thus do not deteriorate the accurate measurement of the longitudinal momenta of the reaction products.

The standard FRS detection equipment was used for the isotopic identification. It consists of two plastic scintillation detectors, two Multiple-Sampling Ionization Chambers (MUSIC) [21] and a system of Multiwire-Proportional Counters (MWPC) [22] located as shown in Fig. 1. In order to cover the full nuclear-charge range, the measurements performed in the present experiments were split into two groups of settings of the FRS. In the light-fragment settings the fragments with charge $Z<30$ were measured, while in the heavy-fragment settings fragments with charge $Z>25$ were detected in order to have a sufficient overlap of the measured cross sec-tions. In the case of light-fragment settings a degrader in the intermediate image plane was used so that only the fragments up to $Z\sim30$ were transmitted through the FRS. Moreover, due to the limited momentum acceptance of the FRS, fragments with a relatively narrow range of $A/Z$ and/or velocities (i.e. magnetic rigidity $B\rho$) are transmitted in a given magnetic field. Thus, around 50 different settings of the magnetic fields had to be used within the light and heavy fragment FRS settings, respectively, in order to measure the reaction products in a broad range of $A/Z$ and/or velocities. The magnetic fields of the first two dipoles were changed in steps of 1.5%, and the fields of the last two dipoles were correspondingly adjusted to keep a selected nuclear charge on the central trajectory. A step of 1.5% was chosen, since it corresponds to half the FRS momentum acceptance, which assures sufficient overlap of the velocity distributions measured in the neighbouring settings. In this way, the full velocity distributions could be reconstructed with a high quality. Since the full momentum (velocity) distributions of all residues are reconstructed from measurements performed in several different magnetic-field settings, see Figure 2, the limited acceptance in momentum does not affect the measured results. For each produced nuclide, the integral of its velocity distribution, normalized to the number of beam particles, is evaluated to determine the production cross section.

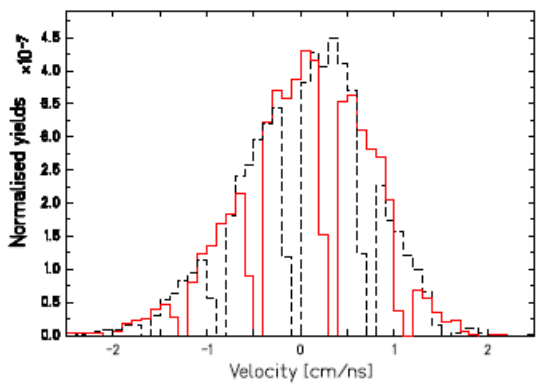

**Figure 2:** (Color online) Velocity distribution of the isotope $^{31}$P measured in the $^{136}$Xe+Pb experiment. Different parts of this distribution measured in neighbouring settings of the magnetic fields are marked by full and dashed histograms.

Due to the limited angular acceptance of the FRS, the transmission of residues having broad angular distributions is reduced and thus also the measured yields. While the heavy residues are produced with rather narrow angular distributions and they are almost fully transmitted through the FRS, the angular distributions of light residues are rather broad, and the angular transmission of these residues may be as low as 10%. As it will be discussed later, the rather low transmission of the light residues may nevertheless be corrected for, using a dedicated transmission calculation [23], which simulates the transmission of every ion species through the magnetic fields of the FRS.

## III Data analysis

The identification of the residues is the key part of the data analysis. In the following, the main steps of the analysis are outlined. A more detailed description of the analysis procedure can be found in ref. [24]. To determine the mass of the final residue, its charge, velocity and magnetic rigidity must be known. The mass identification is then performed according to the relation:

$$\frac{A}{q} = \frac{e}{u}\frac{B\rho}{\beta\gamma c}. \qquad (1)$$

Here, e is the magnitude of the electron charge, u is the atomic mass unit, $\gamma$ represents the relativistic Lorentz factor $\gamma = \sqrt{(1-\beta^2)^{-1}}$ with $\beta = v/c$, c is the speed of light, and $q$ corresponds to the ionic charge of the fragment.

On the passage through various layers of matter in the beam line, the produced fragments may catch or loose electrons. These processes depend on the velocity of the fragments, the velocity of the electrons in their orbits and the material of the layer of matter. Ions with different charge states follow different trajectories in the FRS. This change of trajectory of the ion in the magnetic field following the capture or loss of an electron was used to separate the charge states of the final resi-dues and to remove the majority of incompletely stripped ions from the analysis [25]. Nevertheless, a small fraction of incompletely stripped ions, which cannot be identified this way, still remains and may contaminate the cross sections of completely stripped ions. In the case of both investigated reactions this contribution was well below 1% for all the residues. Thus, in the following only residues with $q=Z$ are considered to enter the above equation.

The velocity of every residue is determined from the measurement of its time-of-flight using the relation $v = l/ToF$, where $l$ is the length of the flight path of the residue. To determine the fragment velocity, its time-of-flight is measured between the scintillation detectors located in the intermediate dispersive and final achromatic image planes F2 and F4 (see Fig. 1), respectively, over the flight path of 36.8 m. The resolution of the ToF determination is given by the time resolution of the scintillation detectors, and is as good as 100 ps (FWHM). This allows determining the velocity parameter $\beta\gamma$ with a relative uncertainty of $2.8 \cdot 10^{-3}$.

Since the time-of-flight measurement is performed over the second stage of the Fragment Separator, the magnetic rigidity $B\rho_{II}$ of every residue in the last two dipoles of the FRS must be determined for use in equation (1). The determination of the magnetic rigidity consists of the measurement of the radius of the fragment trajectory in the second stage of the FRS for a given magnetic field. The value of the magnetic field is measured by Hall probes, and the radius of the trajectory is determined by measuring the fragment positions in two scintillation detectors situated at the final achromatic (F$_4$) and intermediate dispersive (F$_2$) image planes, respectively. The resolution in magnetic rigidity is given by the ion-optical properties of the FRS and by the position resolution of the scintillation detectors, and is as high as $\Delta B\rho/(B\rho) = 5 \cdot 10^{-4}$ (FWHM).

Apart from velocity and magnetic rigidity, also the nuclear charge of every residue must be known in order to perform the mass identification according to equation (1). To measure the charge of the produced fragments, two Multiple-Sampling Ionization Chambers located at F4 behind the scintillation detector were used. In order to obtain the nuclear charge of the residue with high resolution, several corrections must be applied to the measured energy-loss signal. The final, corrected energy loss reads:

$$\Delta E_{corr} = \Delta E_{meas} \cdot f_1(v) \cdot f_2(x) \cdot f_3(T,p) \qquad (2)$$

where $f_1(v)$ is a correction for the velocity dependence of the energy loss, $f_2(x)$ is used to correct for the dependence of the energy-loss signal on the horizontal position in the final image plane, and $f_3(T,p)$ is a correction for the change of pressure and temperature in the ionization chamber. A more detailed description of these corrections may be found in [25].

### III.1 Mass and nuclear-charge identification

Once fully stripped ions are selected, their mass and nuclear charge may be identified. The easiest way to obtain the identification is to display the energy-loss signals measured by the MUSIC versus the $A/Z$ ratio calculated using equation (1).

In Figure 3, the identification plots from the $^{136}$Xe+Pb experiment as measured in the light- and heavy-fragment settings are shown. Each single spot in the figure corresponds to an individual fragment of a given mass $A$ and a nuclear charge $Z$. The high resolution in mass and nuclear charge achieved in this experiment may be seen in the clear separation of the spots corresponding to different nuclides. The mass resolving power, defined by the resolution of the measurements of magnetic rigidity ($B\rho$) and ToF, corresponds to $A/\Delta A \approx 400$ even for the heaviest residues. The resolution of the nuclear charge is as good as $\Delta Z=0.4$ units (FWHM) for all residues. This allows to isotopically identify all the reaction products ranging from the lightest ones up to the heavy projectile. A regular pattern may be observed in the positions of the single nuclides: The fragments to the right from the $A/Z=2$ chain (top panel in Fig.3) correspond successively to isotopes with $N = Z+1$, $Z+2$ etc., while fragments to the left correspond to $N = Z-1$, $Z-2$ etc. This pattern continues in the heavy-fragment settings. The mass and charge identification may now be easily performed using the known $A$ and $Z$ of the particle-unstable nuclei, i.e. $^8$Be, $^9$B, and $^{16}$F and the regular pattern of $N=Z+k$ ($k$ = -2,-1…).

charge distributions as well as of $<N>/Z$ determined in the light and heavy data sets. The isotopes of $Z=3$ were the lightest residues detected and identified in the $^{136}$Xe+Pb experiment.

Similar considerations were applied for the identification of residues from the $^{124}$Xe+Pb experiment, and the corresponding

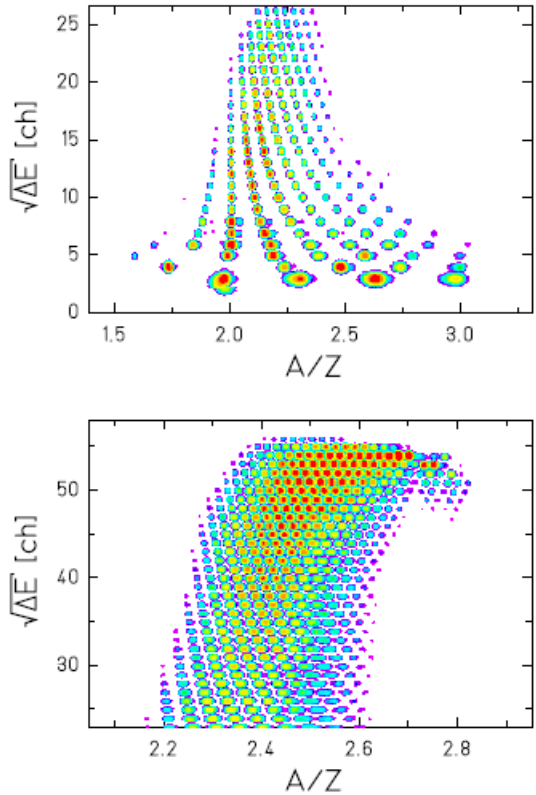

**Figure 3**: (Color online) Identification plots from the $^{136}$Xe+Pb experiment: (top) light-fragment settings; (bottom) heavy-fragment settings. The vertical scale corresponds approximately to the atomic number $Z$.

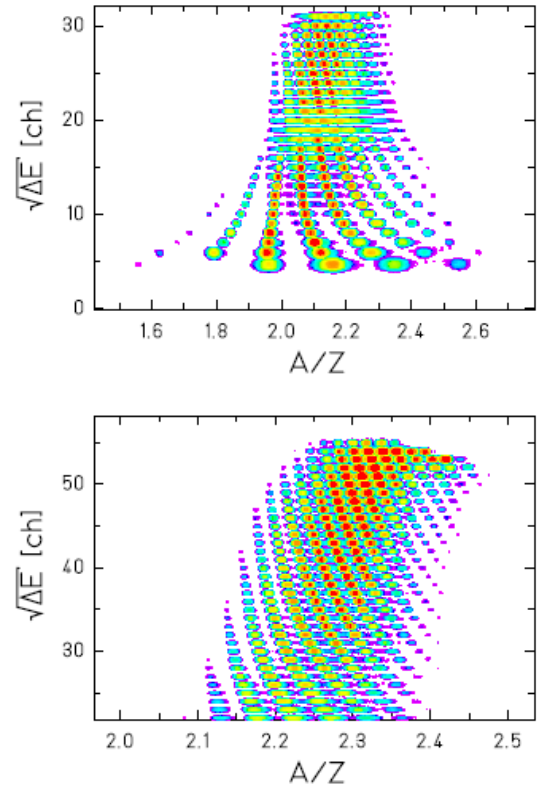

**Figure 4**: (Color online) Identification plots from the $^{124}$Xe+Pb experiment: (top) light-fragment settings; (bottom) heavy-fragment settings. The vertical scale corresponds approximately to the atomic number $Z$.

The regular pattern and the known $A$, $Z$ of the projectile may be used to extend the identification towards heavy residues. In this way, a unique identification both in nuclear charge and mass may be performed. Note that the slight deviation of $A/Z=2$ from a strictly vertical line, observed in Fig. 3 top, does not affect the result of the nuclide identification, after which every fragment is fully specified in $A$ and $Z$ as integer numbers. The validity of the identification of single nuclides is confirmed by a correct 'overlap' of the mass and nuclear identifications plots are shown in Figure 4. In this experiment, the lowest nuclear charge measured was $Z=5$. Also in case of this experiment a high mass and nuclear-charge resolution was achieved as may be seen in the clear separation of the spots corresponding to different nuclides.

From the top part of Figure 4, a poor mass resolution may be observed in the range corresponding to nuclear charges $Z=18-22$. This is a consequence of a malfunction of the constant-fraction discriminator used to obtain the posi-tion information from the arrival time of the signal from one side of the scintillation detector located in the intermediate image plane. Despite this, it was possible to reconstruct the isotopic identification and determine the production cross sections of the single isotopes by means of the time-of-flight measurement for all but the potassium isotopes, which had to be reconstructed using data measured in the heavy-fragment magnetic field settings. The corresponding production cross sections of these isotopes, however, suffer from larger systematic uncertainty.

### III.2 Determination of the production cross sections and velocity distributions

To determine the production cross sections from the measured yields of single nuclides, the following relation is used:

$$\sigma(N,Z) = \frac{Y_{meas}(N,Z)}{N_{Pb} \cdot T(N,Z)} \cdot s(A) \quad (3)$$

Here $N_{Pb}$ is the number of lead-target nuclei per unit area, $Y_{meas}(N,Z)$ stands for the dead-time-corrected production rate per incident projectile, $T(N,Z)$ represents the correction factor for the transmission losses due to the limited angular acceptance of the FRS, and $s(A)$ includes corrections for secondary interactions in the materials in the beam line.

After the identification of $A$ and $Z$, the value of $\beta\gamma$ for each fragment was recalculated from the following equation:

$$\beta\gamma = \frac{e}{c} \cdot \frac{Z \cdot B\rho}{M(A,Z)} \quad (4)$$

where $M(A,Z)$ is the mass of the nucleus $(A,Z)$. In this way, the final resolution in the $\beta\gamma$ measurement is given only by the resolution in $B\rho$, since $A$ and $Z$ are integer numbers after identification, and is, therefore, improved by almost an order of magnitude as compared to the resolution obtained from the TOF measurement. In order to obtain the correct shape of the velocity distribution, the number of counts measured in each magnetic-field setting was normalized to the number of beam ions impinging on the target in this setting. Finally, the velocity distributions from single settings were combined, and the full velocity distribution as illustrated in Figure 2 for $^{31}$P was obtained.

The yields obtained by the integration of these velocity distributions still need to be corrected for the limited angular acceptance of the FRS. For this purpose, the transmission calculation developed in ref. [23] was applied. The model is based on the assumption that momentum distributions have isotropic Gaussian shape around the mean value of the momentum of the emitting source. Using this calculation, the transmission coefficients were determined. The importance of the transmission correction decreases with increasing nuclear charge of the residue, since heavier residues are produced with narrower angular distributions, and the angular acceptance of the FRS is adapted to the emittance of heavy fragmentation products. In both experiments, the transmission $T(N,Z)$ varied between 25% for $Z\sim10$ and reached 100% for $Z\sim40$. The uncertainty of the transmission correction decreases with increasing nuclear charge and varies between 9% for $Z=10$ and values below 1% for $Z>33$.

The correction for secondary reactions in the materials in the FRS beamline (scintillation detector, degrader) was performed using calculations [26,27] based on the Glauber approach. The corrections range from 3% for $A=5$ to 10% for $A=60$ in case of light-fragment settings and from 6% for $A=50$ to 10% for $A=136$ in case of heavy-fragment settings. The higher correction values in the case of light-fragment settings are caused by the thick degrader, which was not used in the heavy-fragment settings. The uncertainty of the secondary-reaction calculation corresponds to app. 10%, which results in a relative systematic uncertainty of production cross sections between 0.3-1% for $A=5-60$ and 0.6-1% for $A=50-136$. The contribution of the secondary reactions in the target material does not exceed 1% for all measured fragments.

# IV Experimental results

In this section the experimental results concerning isotopic, mass and nuclear-charge distributions as well as velocity distributions of the residues formed in the reactions of $^{124,136}$Xe+Pb at 1 $A$ GeV will be presented.

## IV.1 Velocity distributions

The approach, outlined above, to determine the full velocity distributions inside the angular acceptance of the FRS was applied for both measured systems, and examples of some velocity distributions are shown in Figure 5.

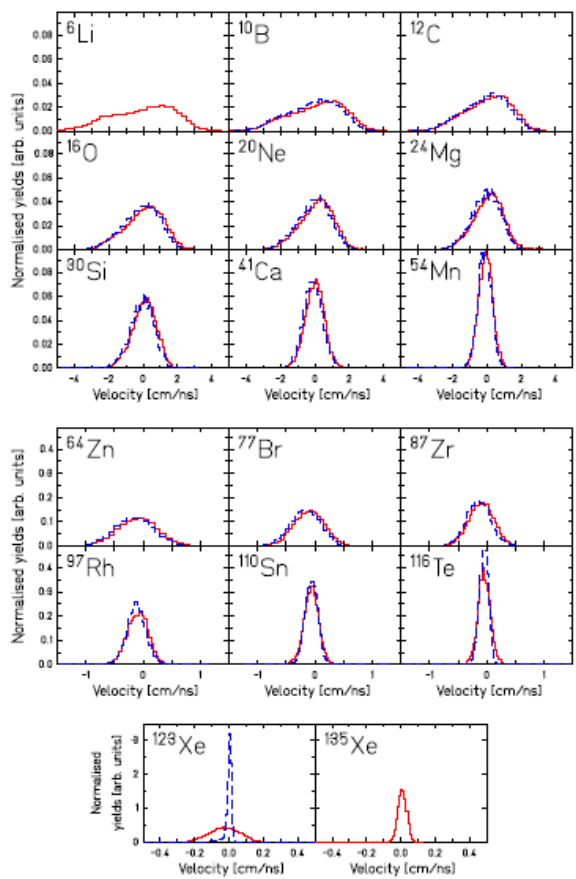

**Figure 5:** (Color online) Measured longitudinal velocity distributions in the projectile frame for several nuclides measured in the $^{136}$Xe+Pb (full red histogram) and the $^{124}$Xe+Pb (dashed blue histogram) experiments. Note the different horizontal scales for different subgroups of data.

For most residues, the longitudinal velocity distributions show Gaussian-like shapes. The tendencies observed in the upper mass range agree with those expected from systematics [9]: The width increases and the mean value decreases with decreasing mass. The width reveals the influence of the Fermi momenta of the abraded nucleons [28] as well as the influence of particle evaporation from the thermally equilibrated system [29], while the mean value reflects the friction experienced by the projectile spectator in the abrasion process. For elements below magnesium, the longitudinal velocity distributions become more complex and deviate from a Gaussian shape. Between neon and carbon, they are asymmetric with a tail to lower velocities. For lithium, a second peak even develops at low velocities. Due to the limited angular acceptance of the FRS, the data for the lightest elements rather correspond to the variation of the invariant cross section along the beam direction [30].

Complex structures in the invariant cross sections have been observed previously for the light residues of the systems $^{56}$Fe + $^1$H, Ti [30] from another experiment at the FRS. In $^{56}$Fe + $^1$H, the lightest residues showed double-humped distributions, typical for binary asymmetric mass splits. In contrast, the system $^{56}$Fe + Ti showed Gaussian-like distributions over the whole mass range. This would be compatible with assuming multifragmentation as the dominant production mechanism for the lightest residues in reactions with the heavier titanium target. These results show that fragments of the same size can be produced by different reaction mechanisms. This observation suggests that the present data may be interpreted as a manifestation of different mechanisms resulting in the production of a given fragment in the same system. However, a more detailed discussion on this subject, which requires performing dedicated model calculations with a suitable nuclear-reaction code, is beyond the scope of the present paper, and will be the subject of a forthcoming paper.

In the following it is assumed that the fluctuations of the velocity distributions of the residues with atomic number larger than 9 are isotropic in space around the average emitting-source velocity. Under this condition, the losses due to the limited angular acceptance of the FRS can be estimated using the algorithm of ref. [23]. For the lighter residues, only the directly measured production cross sections inside the angular acceptance of the FRS will be given.

## IV.2 Angular-acceptance-integrated production cross sections

In order to provide the full isotopic distributions over a broad range of elements, the production cross sections over several orders of magnitude had to be measured. An overview of the complete dataset is shown in Figure 6. The production cross sections measured in both experiments extend over a range of $\approx 1\mu b$ to 2b with the production cross sections for the isotopes of a single element spanning in most cases over three orders of magnitude.

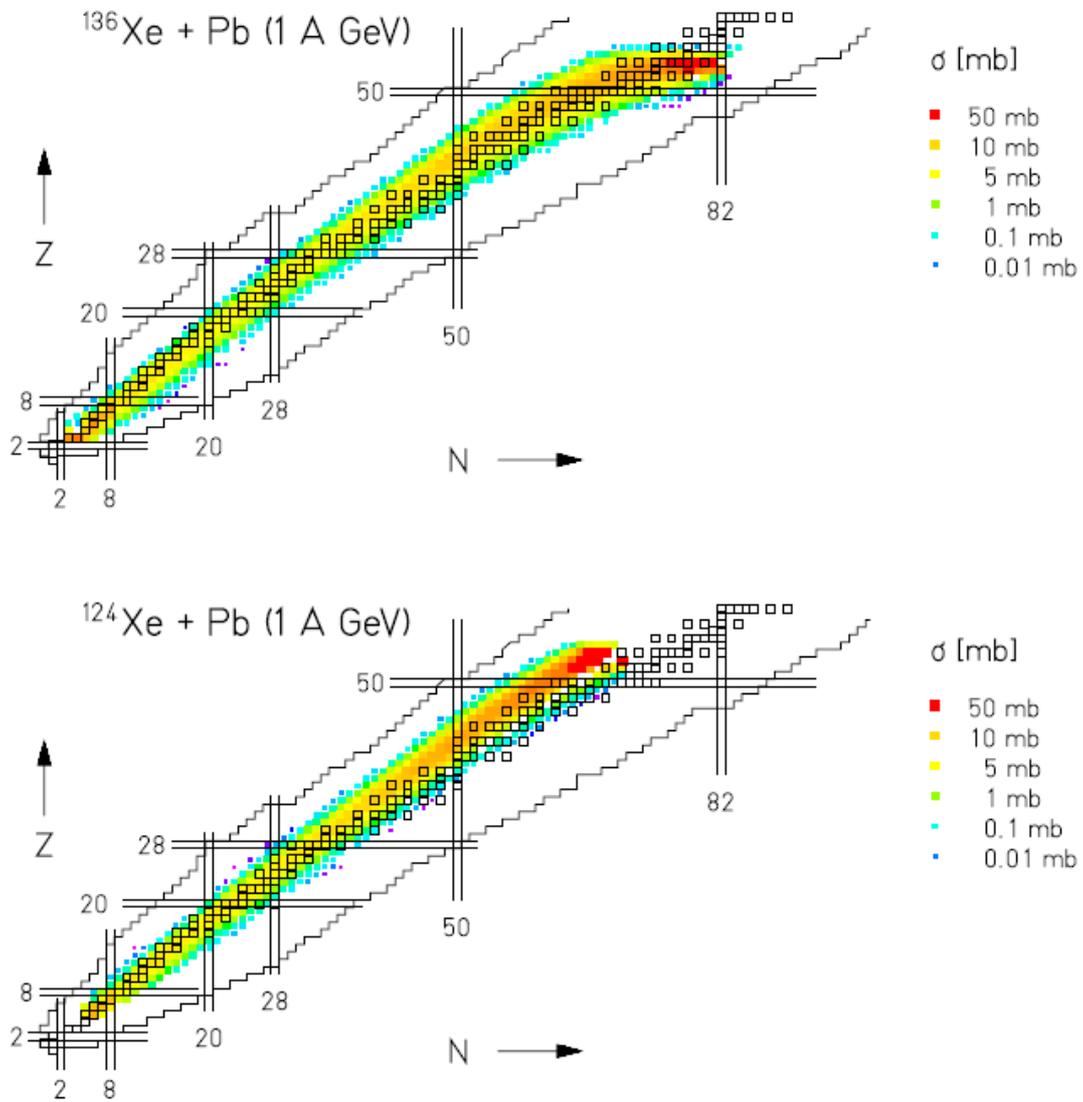

**Figure 6:** (Color online) The angular-acceptance-integrated production cross sections for isotopes of elements $Z$=5-55 and $Z$=3-56 measured in the $^{124}$Xe+Pb and $^{136}$Xe+Pb experiments, respectively, presented on the chart of nuclides. Missing isotopes close to the projectile in the $^{124}$Xe+Pb chart are due to more restrictive slit settings in this experiment (see the text).

The angular-acceptance-integrated production cross sections (not corrected for the limited FRS acceptance affecting mainly lighter residues) were determined for all nuclides measured in the two experiments which correspond to isotopes of elements $Z$=5-55 and $Z$=3-56 in case of the $^{124}$Xe+Pb and the $^{136}$Xe+Pb experiments, respectively. The angular-acceptance-integrated cross section $\sigma_{acc}$ may be expressed as:

$$\sigma_{acc} = \int_0^{\alpha_{FRS}} \frac{d\sigma}{d\alpha} d\alpha \qquad (4)$$

where $\alpha_{FRS}$ denotes the FRS acceptance of 15 mrad around the beam axis. The interest in the angular-acceptance integrated cross sections lies in the fact that they provide directly measured quantities independent of the assumptions on the velocity distributions in the full velocity space needed to model the transmission of fragments through the FRS and on the transmission calculation itself. Since the velocity distributions of the fragments from both reactions are similar, the angular-acceptance-integrated isotopic distributions are well suited for the relative comparison of the products of the two reactions. The numerical values of the transmission-corrected production cross sections for both experiments are listed in Annex A.

The angular-acceptance-integrated isotopic distributions from both experiments measured in the nuclear-charge range $Z$=5-55 ($^{124}$Xe+Pb) and $Z$=3-56 ($^{136}$Xe+Pb) are compared in Figure 7. Due to the thresholds of the electronics, some losses may be expected in case of $Z$=3 isotopes measured in the $^{136}$Xe+Pb experiment, and the corresponding cross sections introduced in Figure 7 should be considered as lower limits only. Please, note that not all the isotopes visible in the identification pattern in Figure 3 and Figure 4 may be found in Figure 7. In case of the lightest residues this is due to their broad velocity distributions, which were not fully measured for all the detected isotopes, and thus the cross section could not be properly determined. Velocity distributions of several heavy isotopes (see table A.1) in case of $^{124}$Xe+Pb were severely cut by slits, which were inserted to protect the detectors from the most intense charge states of the primary beam with zero, one and two electrons, and thus their cross sections could not be recovered. Some heavy neutron-rich isotopes from the $^{136}$Xe experiment were affected as well (48≤$Z$≤52), resulting in the apparent deviation of the corresponding cross sections from a smooth trend.

Several interesting observations can be made by comparing the shapes of the isotopic distributions produced in the reactions with the two projectiles largely differing in the initial neutron-to-proton ratio $N/Z$ ($^{124}$Xe with $N/Z$=1.30 and $^{136}$Xe with $N/Z$=1.52). A slightly enhanced production of more neutron-rich isotopes is observed in the isotopic distributions of the lightest elements ($Z \approx$ 5-9) measured in the fragmentation of the more neutron-rich $^{136}$Xe projectile. This enhancement of cross sections for neutron-rich isotopes is replaced by a shift of the isotopic distributions towards more neutron-rich isotopes for elements with charge above $Z \approx$ 10. A clear difference between the positions of the maxima of the isotopic distributions from fragmentation of $^{124}$Xe and $^{136}$Xe projectiles may be observed, which increases with increasing nuclear charge. The largest difference is observed for the elements in the vicinity of the projectile, since here a pronounced memory on the initial isotopic composition is preserved due to rather low excitation energies acquired in the collision. With decreasing nuclear charge this memory is considerably reduced due to higher excitation energies introduced in the collision and, thus, a longer deexcitation process. Nevertheless, already from this comparison it is obvious that the memory on the initial $N/Z$ is preserved in the whole nuclear-charge range despite the influence of the evaporation process.

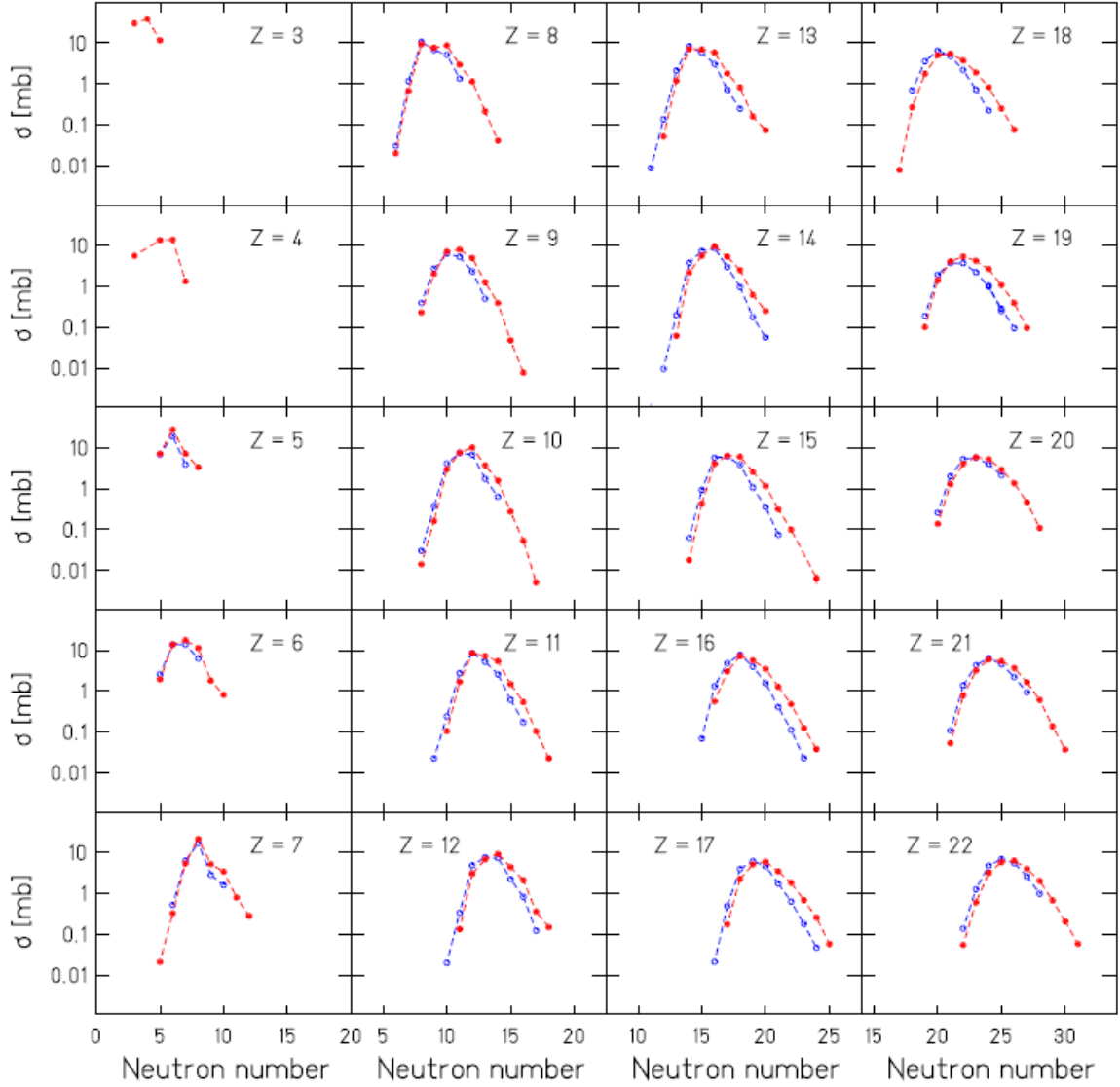

**Figure 7:** (Color online) Angular-acceptance-integrated isotopic cross sections measured in $^{136}$Xe+Pb (full red symbols) and $^{124}$Xe+Pb (open blue symbols). The dashed lines serve to guide the eye. Statistical error bars are smaller than the size of the symbols.

Apart from the observation that a more neutron-rich projectile results in more neutron-rich final residues, the isotopic distributions from the $^{136}$Xe projectile appear to be broader, which is especially pronounced for elements in the vicinity of the projectile. The isotopic distribution of primary fragments produced by the initial collision transforms into the distribution of final fragments through the emission of neutrons, protons and complex clusters. Due to the higher neutron excess of the $^{136}$Xe projectile, the evaporation from elements close to the projectile may be viewed as progressing close to a horizontal line of constant $Z$ in the chart of the nuclides, populating thus a rather broad range of isotopes of the same element. On the contrary, in case of the less neutron-rich projectile $^{124}$Xe the competition between neutron and proton evaporation depopulates a given isotopic chain in favor of producing isotopes of lower elements. As a consequence, only a rather narrow range of isotopes of the same element is populated in the evaporation in case of this less neutron-rich projectile. Narrower isotopic distributions in the vicinity of the projectile may, thus, be expected for $^{124}$Xe.

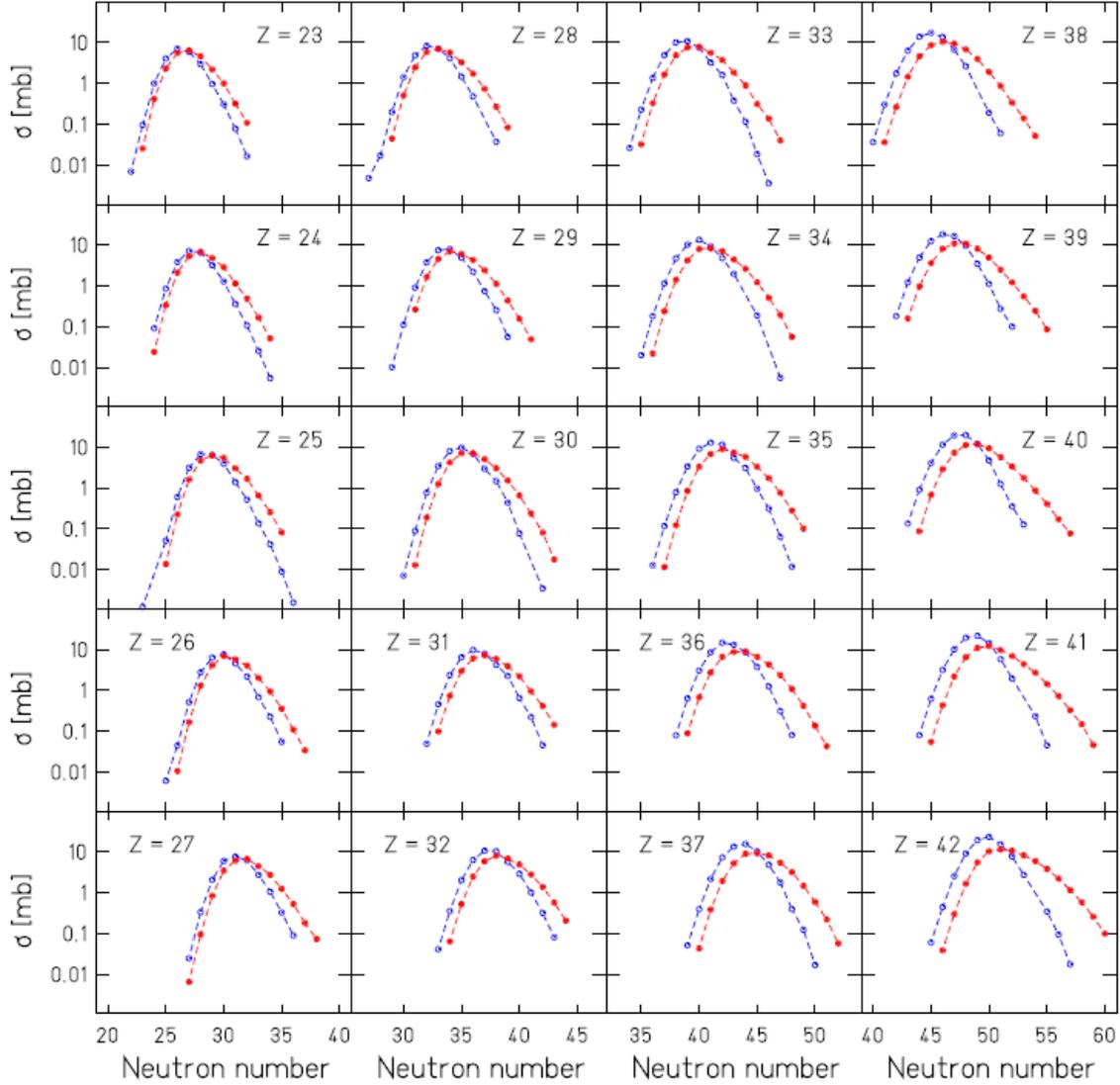

**Figure 7**: Continuation.

In addition to the trends in the mean values and widths, another interesting feature may be observed in the final isotopic distributions from both experiments. In the nuclear-charge range $Z \approx 5\text{-}15$ a staggering in the cross sections of neighboring isotopes can be seen. This staggering is a manifestation of an even-odd effect resulting from the condensation process of heated nuclear matter while cooling down in the evaporation process [31]. As such, it can be considered as the manifestation of the passage from the normal liquid phase of the nucleus to its superfluid phase. The even-odd staggering disappears around $Z\sim15$ due to the increasing competition of the γ-emission as a consequence of increasing level density below the particle emission threshold. In ref. [31] a quantitative discussion of the even-odd staggering in the production cross sections is given. It is remarkable in the present data that also for the heavy nuclides in the vicinity of the projectile an even-odd staggering in the production cross sections is present ($Z$=53). This is a new observation absent in the previous FRS experiments with heavier projectiles (Pb+Cu, Pb+$^{1,2}$H, Au+$^{1}$H) [32,24,33,34].

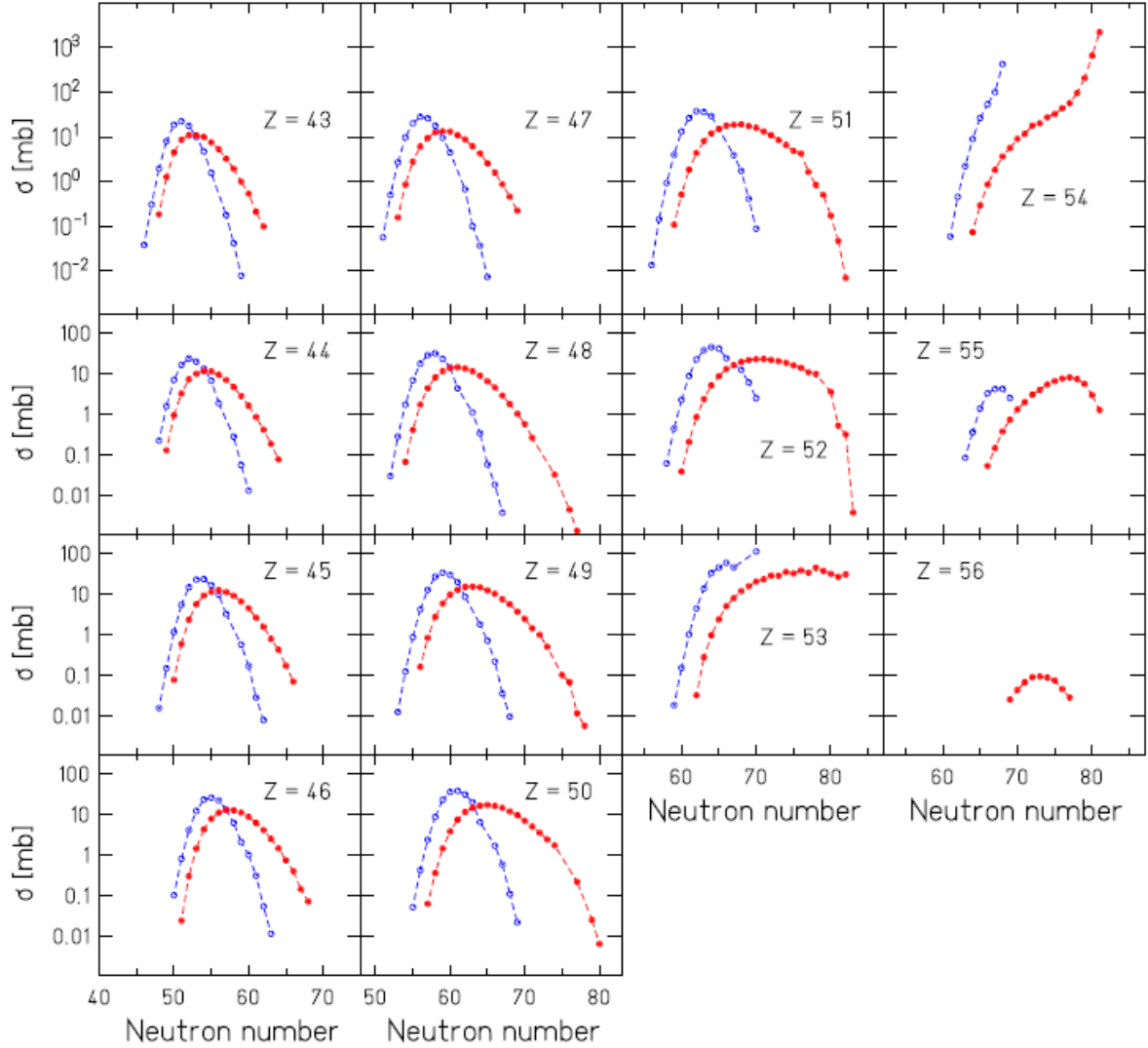

**Figure 7**: Continuation.

In both experiments also the charge-pickup reactions were measured (isotopes with $Z>Z_{projectile}$). While in case of the less neutron-rich projectile only single charge-pickup channels are observed ($Z=55$), in case of $^{136}$Xe also double charge-pickup residues were detected ($Z=56$). Charge-pickup reactions proceed [35,36] either through a quasi-elastic collision between a proton and a neutron of the target and the projectile nucleus, respectively, where the proton replaces the neutron inside the projectile-like fragment, or through the excitation of a projectile or a target nucleon into the $\Delta(1232)$-resonance state and its subsequent decay. In case of the more neutron-rich projectile the primary projectile-like fragment after the charge-pickup reaction is still rather neutron-rich so that the neutron emission dominates in the evaporation process. As a consequence, the isotopes of the same nuclear charge as the excited projectile-like fragment are predominantly produced. Again, the competition between the neutron and proton emission in case of the primary projectile-like fragment from the less neutron-rich $^{124}$Xe may be the reason, why the double charge-pickup residues (isotopes of $Z=56$ element) are not observed in case of this projectile.

## IV.3 Mass and nuclear-charge distributions

To obtain the full production cross sections, the correction for the limited angular transmission of single isotopes through the FRS must be applied. The correction for the limited angular transmission through the FRS was performed only for isotopes of elements with $Z \geq 10$ in both experiments. The velocity distributions for the lighter elements reveal structures presumably originating from the overlap of contributions from different reaction mechanisms, which makes the determination of the transmission correction for these isotopes rather difficult. Therefore, the production cross sections in the following are restricted to the final residues in the nuclear-charge range $Z \geq 10$, where the transmission correction was evaluated under the assumption of isotropic Gaussian-shaped momentum distributions around the mean value of the emitting-source momentum.

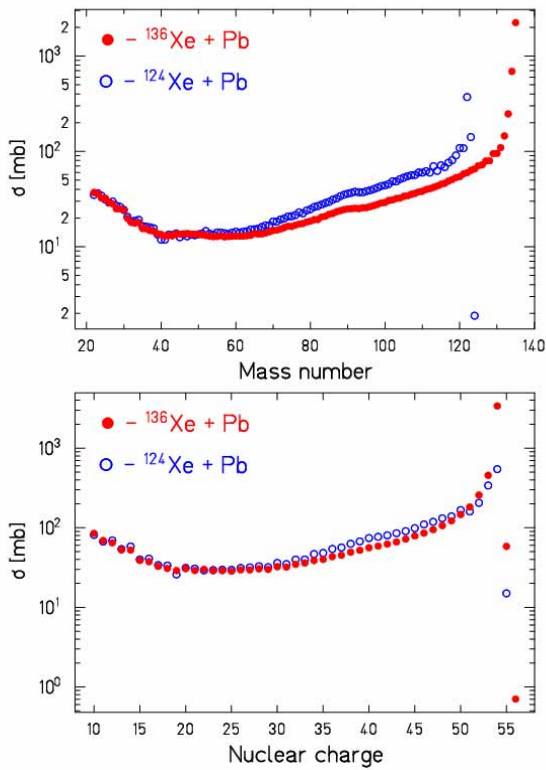

**Figure 8**: (Color online) Mass (top) and nuclear-charge (bottom) distributions measured in the $^{136}$Xe+Pb (full red symbols) and $^{124}$Xe+Pb (open blue symbols) experiments. Statistical error bars are smaller than the size of the symbols.

The mass and nuclear-charge distributions determined from the production cross sections measured in both experiments and corrected for the FRS angular transmission are presented in Figure 8. Systematic uncertainties are the same as those of the nuclide yields in the corresponding element respectively mass range. The slight depletion observed in the mass distribution for the reaction $^{124}$Xe+Pb in the mass range $A \sim 40$ is a consequence of the fact that the isotopes of the element $Z=19$ could not be reconstructed from the light-fragment settings. Rather they had to be extracted from the heavy-fragment settings where they passed the FRS rather close to its borders and consequently the yields of these isotopes were slightly cut by the FRS acceptance. Similarly, a slightly lower value of the total elemental cross section of this element is observed in the charge distribution.

Overall, rather similar trends in case of both projectiles are seen, characterized by steeply decreasing cross sections of the heavy residues with decreasing mass and nuclear charge, followed by a plateau of rather constant cross sections below $A \sim 65$ ($Z \sim 30$) and an exponential increase of the cross sections of the light fragments. In the earlier investigations of the mass and nuclear-charge distributions, the shape of the nuclear charge or mass distributions was observed to evolve from two distinct regions corresponding to heavy residues close to the projectile (or target) and to light fragments produced in evaporation, towards the U-shape nuclear charge and mass distributions [37]. The former case is typical for the sequential evaporation from a moderately excited nuclear source, while the U-shape distributions were usually observed in case of collisions where considerably higher excitation energies were introduced, leading eventually to the simultaneous break-up of the highly excited nuclear source [38]. Exploring the mass or nuclear-charge distributions measured in the present experiments, it is observed that the cross sections of residues in the plateau regions are approximately only a factor of 2-3 lower than the cross sections of the lightest fragments observed. This moderate change of the cross section as a function of mass and nuclear charge is more similar to trends of the mass and nuclear-charge distributions measured in reactions with considerably high excitation energies introduced in the collision, which may proceed through a break-up stage.

## IV.4 Mean *N*-over-*Z* ratio

Already from the comparison of the measured isotopic distributions, the enhanced production of more neutron-rich isotopes in the $^{136}$Xe+Pb reaction could be observed. This trend suggests a dependence of the final isotopic composition on the *N/Z* of the projectile, and may be studied in more detail if the mean values of the isotopic distributions from both experiments are compared. For this purpose, the mean *N*-over-*Z* ratio (<*N*>/*Z*) is determined from each isotopic distribution allowing a direct comparison with the *N/Z* of the two projectiles. Figure 9 shows the <*N*>/*Z* of the final residues measured in both experiments as a function of the nuclear charge.

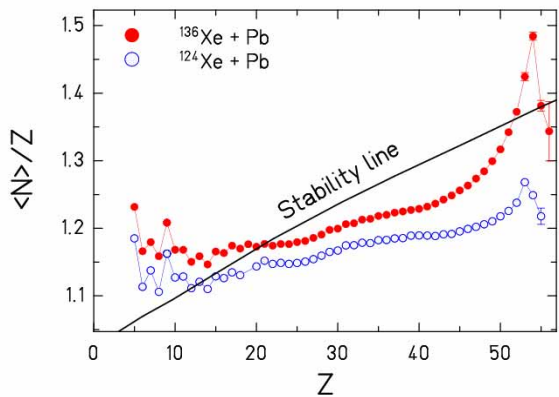

**Figure 9:** (Color online) Comparison of the <*N*>/*Z* of final residues determined from the isotopic distributions of the reactions $^{136}$Xe(*N/Z*=1.52)+Pb (full red symbols) and $^{124}$Xe(*N/Z*=1.30)+Pb (open blue symbols) at 1 *A* GeV. Values for *Z* < 10 are derived from angular-acceptance-integrated cross sections. The full line represents the <*N*>/*Z* of the stable isotopes [39]. Statistical error bars are not shown when they are smaller than the size of the symbols. Systematic uncertainties are negligible.

The data are compared with the stability line obtained from ref. [39]. Three different regions may be identified in the data:

(1) For the nuclei in the vicinity of the projectile (app. 50≤*Z*≤54) a rather steep decrease of the <*N*>/*Z* values is observed with decreasing nuclear charge for the case of the $^{136}$Xe projectile. This decrease appears to be less steep in case of the $^{124}$Xe projectile. These observations may be understood as a consequence of very peripheral collisions producing the final residues in the vicinity of the projectile, where rather low excitation energies are introduced. In case of the $^{136}$Xe projectile the neutron emission dominates the evaporation process for the primary fragments in this nuclear-charge range. This prevalent neutron emission strongly affects the *N/Z* of the residues, which results in the steep trend observed in the figure. On the contrary, in case of the less neutron-rich projectile $^{124}$Xe the emission of neutrons and protons may compete in the evaporation process already at rather low initial excitation energies, which makes the change of the *N/Z* of residues close to this projectile less steep.

(2) In the region below *Z* ≈ 50 the transition to a smoother dependence of the <*N*>/*Z* on the nuclear charge occurs. This is a consequence of higher excitation energies introduced in the collision, and, thus, the competing emission of neutrons, protons, and eventually more complex clusters during the evaporation. Nevertheless, despite the decreasing nuclear charge (i.e. increasing excitation energy acquired in the collision), the final residues from the $^{136}$Xe projectile remain to be more neutron-rich on average as compared to the residues from $^{124}$Xe in the whole nuclear-charge range. This was already observed in the relative shift of the isotopic distributions from the two projectiles. This observation is particularly interesting, since as discussed in refs. [40,41], the isotopic composition of the final residues after a long evaporation process is expected to gradually approach the region of equilibrium neutron- and proton-emission probabilities, known as the evaporation-attractor line (EAL) or the residue corridor [40,41]. Thus, the isotopic composition of the final residues far from the projectile should no longer depend on the *N/Z* of the initial system. The extent to which the corridor appears to be 'attractive' to the final residues depends, however, on the neutron or proton excess of the initial system, which may, due to the emission of more complex clusters in the evaporation, even prevent the final residues from reaching the residue corridor at all [41]. At the same time, the highly excited nuclear source can undergo a simultaneous break-up process, which may affect the isotopic composition as well as the excitation energy of the produced fragments that enter into the evaporation process. As a consequence, the final residues may be prevented from ever reaching the residue corridor [12]. The difference in the final <N>/Z of the final residues is an interesting observation, which may be used to investigate the relative importance of different reaction mechanisms leading to the final isotopic composition. This will be the subject of a forthcoming paper.

(3) The <N>/Z of the lightest nuclei (below Z ≈ 15) strongly staggers as a function of Z. This nuclear-charge range coincides with the region where the even-odd effect is particularly pronounced in the isotopic distributions, as discussed earlier. It is the enhancement of the production of the even-N stable isotopes (mostly N=Z) for even nuclear charge, which is responsible for the shift of the mean values of the even-Z isotopic distributions towards the less neutron-rich isotopes. This shift is reflected in the lower values of the <N>/Z of the final residues with even nuclear charge.

Overall, the <N>/Z provides a rich experimental information, which upon comparison with nuclear-reaction codes can help to extract additional information on the properties of highly excited nuclear systems [12].

## V Summary

Within the current paper, the velocity distributions and cross sections of residues produced in the interactions of $^{124}$Xe(N/Z=1.30) and $^{136}$Xe(N/Z=1.52) projectiles with a lead target were investigated. Both experiments were performed at the high-resolution magnetic spectrometer, the Fragment Separator at GSI, which allows for the identification of the reaction residues in the complete mass and nuclear-charge range. More than 1100 nuclides were measured in both experiments, covering the isotopes of Z=3-56 elements in case of the $^{136}$Xe+Pb reaction and Z=5-55 elements in case of the $^{124}$Xe+Pb system.

The angular-acceptance-integrated cross sections were determined for all isotopes in this nuclear-charge range. The transmission-corrected production cross sections, which correspond to the full production of single isotopes, were determined for isotopes in the nuclear-charge range Z=10-55(56) for the $^{124}$Xe and $^{136}$Xe projectile, respectively. The production cross sections measured in each experiment range over several orders of magnitude from 1μb to 2b with a relative uncertainty corresponding to 8-15% in most cases. The longitudinal momenta of the residues were measured with a relative uncertainty of 5·10$^{-4}$, which allows to investigate the mechanisms responsible for fragment formations.

The measured isotopic distributions reveal an enhancement of cross sections for neutron-rich isotopes of the lightest elements produced in the reaction $^{136}$Xe+Pb as compared to the reaction $^{124}$Xe+Pb. This is then gradually replaced by a shift of the mean values of the isotopic distributions in the reaction $^{136}$Xe+Pb towards the more neutron-rich side for elements with nuclear charge above Z ≈ 10. This observation reveals a clear memory on the N/Z of the projectile being preserved over the whole range of the nuclear charge despite the influence of the evaporation process, and thus the isotopic distributions may be used to investigate the relative importance of different reaction mechanisms leading to the final isotopic composition. Furthermore, these data are of importance for studying the isospin effect in the symmetry energy and for studying the effect of the evaporation on the observables which are related to the EOS.

## Acknowledgements

The support provided during the experiment by the technical staff of the Fragment Separator Karl-Heinz Behr, Adolf Brünle and Karlheinz Burkard as well as the help of Nikolaus Kurz in preparing the data acquisition is gratefully acknowledged.

# Annex A

## Compilation of the cross sections measured in the $^{136}$Xe+Pb and $^{124}$Xe+Pb reactions

The transmission-corrected production cross sections measured in the two experiments, which were analyzed within this work, are summarized in table A.1. The corresponding absolute uncertainties (values include both, statistical and systematical uncertainties) are indicated.

**Table A.1:** The full list of the production cross sections of isotopes of Z=10-55(56) elements measured in the $^{136}$Xe+Pb (left column) and $^{124}$Xe+Pb (right column) experiments. The absolute errors include statistical and systematical uncertainties.

| | $^{136}$Xe+Pb 1 $A$ GeV | | | | $^{124}$Xe+Pb 1 $A$ GeV | | |
|---|---|---|---|---|---|---|---|
| Z | N | σ [mb] | Δσ [mb] | Z | N | σ [mb] | Δσ [mb] |
| 10 | 8  | 0.060 | 0.008 | 10 | 8  | 0.146 | 0.015 |
| 10 | 9  | 0.630 | 0.073 | 10 | 9  | 1.68  | 0.17  |
| 10 | 10 | 10.7  | 1.2   | 10 | 10 | 17.3  | 1.7   |
| 10 | 11 | 25.6  | 2.9   | 10 | 11 | 29.1  | 2.9   |
| 10 | 12 | 31.7  | 3.6   | 10 | 12 | 25.0  | 2.5   |
| 10 | 13 | 10.9  | 1.2   | 10 | 13 | 6.04  | 0.60  |
| 10 | 14 | 4.33  | 0.49  | 10 | 14 | 2.07  | 0.21  |
| 10 | 15 | 0.706 | 0.083 |    |    |       |       |
| 10 | 16 | 0.130 | 0.016 | 11 | 9  | 0.092 | 0.010 |
| 10 | 17 | 0.012 | 0.002 | 11 | 10 | 0.921 | 0.092 |
|    |    |       |       | 11 | 11 | 9.87  | 0.98  |
| 11 | 10 | 0.349 | 0.040 | 11 | 12 | 29.3  | 2.9   |
| 11 | 11 | 5.27  | 0.59  | 11 | 13 | 16.7  | 1.7   |
| 11 | 12 | 24.7  | 2.8   | 11 | 14 | 7.90  | 0.79  |
| 11 | 13 | 19.7  | 2.2   | 11 | 15 | 1.79  | 0.18  |
| 11 | 14 | 13.9  | 1.6   | 11 | 16 | 0.477 | 0.048 |
| 11 | 15 | 3.59  | 0.40  |    |    |       |       |
| 11 | 16 | 1.24  | 0.14  | 12 | 10 | 0.071 | 0.008 |
| 11 | 17 | 0.227 | 0.027 | 12 | 11 | 1.122 | 0.112 |
| 11 | 18 | 0.047 | 0.007 | 12 | 12 | 15.4  | 1.5   |
|    |    |       |       | 12 | 13 | 23.0  | 2.2   |
| 12 | 11 | 0.390 | 0.044 | 12 | 14 | 21.3  | 2.1   |
| 12 | 12 | 8.38  | 0.93  | 12 | 15 | 6.12  | 0.61  |
| 12 | 13 | 17.5  | 1.9   | 12 | 16 | 2.17  | 0.22  |
| 12 | 14 | 22.6  | 2.5   | 12 | 17 | 0.307 | 0.031 |
| 12 | 15 | 10.2  | 1.1   |    |    |       |       |
| 12 | 16 | 4.61  | 0.51  | 13 | 11 | 0.027 | 0.003 |
| 12 | 17 | 0.752 | 0.085 | 13 | 12 | 0.407 | 0.041 |
| 12 | 18 | 0.298 | 0.035 | 13 | 13 | 5.98  | 0.59  |
|    |    |       |       | 13 | 14 | 22.6  | 2.3   |
| 13 | 12 | 0.133 | 0.016 | 13 | 15 | 15.0  | 1.5   |
| 13 | 13 | 2.90  | 0.31  | 13 | 16 | 7.63  | 0.76  |
| 13 | 14 | 16.6  | 1.8   | 13 | 17 | 1.71  | 0.17  |
| 13 | 15 | 15.3  | 1.6   | 13 | 18 | 0.569 | 0.057 |
| 13 | 16 | 12.3  | 1.3   |    |    |       |       |
| 13 | 17 | 3.62  | 0.39  | 14 | 12 | 0.027 | 0.003 |
| 13 | 18 | 1.56  | 0.17  | 14 | 13 | 0.529 | 0.053 |
| 13 | 19 | 0.294 | 0.035 | 14 | 14 | 9.73  | 0.97  |
| 13 | 20 | 0.131 | 0.018 | 14 | 15 | 18.1  | 1.8   |
|    |    |       |       | 14 | 16 | 20.4  | 2.0   |
| 14 | 13 | 0.143 | 0.016 | 14 | 17 | 6.77  | 0.67  |
| 14 | 14 | 4.81  | 0.50  | 14 | 18 | 2.11  | 0.21  |
| 14 | 15 | 11.8  | 1.2   | 14 | 19 | 0.377 | 0.038 |
| 14 | 16 | 19.3  | 2.0   | 14 | 20 | 0.116 | 0.012 |
| 14 | 17 | 10.2  | 1.1   |    |    |       |       |
| 14 | 18 | 4.64  | 0.49  | 15 | 14 | 0.152 | 0.016 |
| 14 | 19 | 1.10  | 0.12  | 15 | 15 | 2.19  | 0.22  |
| 14 | 20 | 0.432 | 0.048 | 15 | 16 | 13.0  | 1.3   |
|    |    |       |       | 15 | 17 | 13.4  | 1.3   |
| 15 | 14 | 0.038 | 0.005 | 15 | 18 | 8.12  | 0.81  |
| 15 | 15 | 0.875 | 0.089 | 15 | 19 | 2.19  | 0.22  |
| 15 | 16 | 7.97  | 0.81  | 15 | 20 | 0.710 | 0.071 |
| 15 | 17 | 12.0  | 1.2   | 15 | 21 | 0.144 | 0.015 |
| 15 | 18 | 10.9  | 1.1   |    |    |       |       |
| 15 | 19 | 4.51  | 0.46  | 16 | 15 | 0.156 | 0.016 |
| 15 | 20 | 1.96  | 0.20  | 16 | 16 | 2.87  | 0.29  |
| 15 | 21 | 0.513 | 0.054 | 16 | 17 | 10.2  | 1.0   |
| 15 | 22 | 0.158 | 0.018 | 16 | 18 | 15.8  | 1.6   |
|    |    |       |       | 16 | 19 | 7.88  | 0.78  |
| 16 | 16 | 1.06  | 0.16  | 16 | 20 | 3.00  | 0.30  |
| 16 | 17 | 5.50  | 0.81  | 16 | 21 | 0.765 | 0.076 |
| 16 | 18 | 12.7  | 1.9   | 16 | 22 | 0.205 | 0.021 |
| 16 | 19 | 9.62  | 1.43  | 16 | 23 | 0.041 | 0.004 |
| 16 | 20 | 5.78  | 0.86  |    |    |       |       |
| 16 | 21 | 2.04  | 0.30  | 17 | 16 | 0.044 | 0.005 |
| 16 | 22 | 0.753 | 0.113 | 17 | 17 | 1.00  | 0.10  |
| 16 | 23 | 0.190 | 0.029 | 17 | 18 | 7.84  | 0.78  |
| 16 | 24 | 0.057 | 0.010 | 17 | 19 | 11.8  | 1.2   |
|    |    |       |       | 17 | 20 | 8.62  | 0.86  |
| 17 | 17 | 0.317 | 0.052 | 17 | 21 | 3.15  | 0.31  |
| 17 | 18 | 3.94  | 0.64  | 17 | 22 | 1.12  | 0.11  |
| 17 | 19 | 8.71  | 1.41  | 17 | 23 | 0.309 | 0.031 |
| 17 | 20 | 9.85  | 1.60  | 17 | 24 | 0.079 | 0.008 |
| 17 | 21 | 5.54  | 0.90  |    |    |       |       |
| 17 | 22 | 2.85  | 0.46  | 18 | 18 | 1.33  | 0.15  |
| 17 | 23 | 1.04  | 0.17  | 18 | 19 | 6.60  | 0.74  |
| 17 | 24 | 0.381 | 0.063 | 18 | 20 | 11.8  | 1.3   |
| 17 | 25 | 0.084 | 0.015 | 18 | 21 | 8.39  | 0.93  |
|    |    |       |       | 18 | 22 | 3.78  | 0.42  |
| 18 | 17 | 0.014 | 0.002 | 18 | 23 | 1.22  | 0.13  |
| 18 | 18 | 0.456 | 0.057 | 18 | 24 | 0.366 | 0.041 |
| 18 | 19 | 2.93  | 0.37  |    |    |       |       |
| 18 | 20 | 8.27  | 1.03  | 19 | 19 | 0.388 | 0.055 |
| 18 | 21 | 8.71  | 1.09  | 19 | 20 | 3.88  | 0.55  |
| 18 | 22 | 5.85  | 0.73  | 19 | 21 | 7.32  | 1.03  |
| 18 | 23 | 2.89  | 0.36  | 19 | 22 | 7.12  | 1.01  |
| 18 | 24 | 1.24  | 0.16  | 19 | 23 | 4.14  | 0.58  |
| 18 | 25 | 0.366 | 0.047 | 19 | 24 | 1.77  | 0.25  |
| 18 | 26 | 0.109 | 0.018 | 19 | 25 | 0.512 | 0.072 |
| | | | | 19 | 26 | 0.167 | 0.024 |
| 19 | 19 | 0.161 | 0.019 | | | | |
| 19 | 20 | 2.19 | 0.25 | 20 | 20 | 0.444 | 0.050 |
| 19 | 21 | 6.26 | 0.72 | 20 | 21 | 3.39 | 0.38 |
| 19 | 22 | 8.07 | 0.93 | 20 | 22 | 8.68 | 0.97 |
| 19 | 23 | 6.25 | 0.72 | 20 | 23 | 9.26 | 1.03 |
| 19 | 24 | 3.84 | 0.44 | 20 | 24 | 6.24 | 0.69 |
| 19 | 25 | 1.52 | 0.17 | 20 | 25 | 3.27 | 0.37 |
| 19 | 26 | 0.549 | 0.064 | | | | |
| 19 | 27 | 0.131 | 0.018 | 21 | 21 | 0.179 | 0.021 |
| | | | | 21 | 22 | 2.20 | 0.25 |
| 20 | 20 | 0.209 | 0.019 | 21 | 23 | 6.80 | 0.76 |
| 20 | 21 | 1.92 | 0.17 | 21 | 24 | 9.93 | 1.11 |
| 20 | 22 | 5.97 | 0.54 | 21 | 25 | 6.91 | 0.77 |
| 20 | 23 | 8.46 | 0.77 | 21 | 26 | 3.30 | 0.37 |
| 20 | 24 | 7.51 | 0.68 | 21 | 27 | 1.38 | 0.15 |
| 20 | 25 | 4.03 | 0.37 | | | | |
| 20 | 26 | 1.86 | 0.17 | 22 | 22 | 0.22 | 0.02 |
| 20 | 27 | 0.620 | 0.057 | 22 | 23 | 1.98 | 0.22 |
| 20 | 28 | 0.144 | 0.014 | 22 | 24 | 7.36 | 0.82 |
| | | | | 22 | 25 | 10.4 | 1.15 |
| 21 | 21 | 0.077 | 0.007 | 22 | 26 | 8.15 | 0.91 |
| 21 | 22 | 1.11 | 0.09 | 22 | 27 | 3.79 | 0.42 |
| 21 | 23 | 4.44 | 0.39 | 22 | 28 | 1.42 | 0.16 |
| 21 | 24 | 8.16 | 0.73 | | | | |
| 21 | 25 | 7.27 | 0.65 | 23 | 22 | 0.011 | 0.001 |
| 21 | 26 | 4.87 | 0.43 | 23 | 23 | 0.150 | 0.015 |
| 21 | 27 | 2.14 | 0.19 | 23 | 24 | 1.53 | 0.15 |
| 21 | 28 | 0.777 | 0.069 | 23 | 25 | 6.08 | 0.61 |
| 21 | 29 | 0.174 | 0.017 | 23 | 26 | 10.4 | 1.0 |
| 21 | 30 | 0.046 | 0.006 | 23 | 27 | 8.63 | 0.86 |
| | | | | 23 | 28 | 4.25 | 0.42 |
| 22 | 22 | 0.076 | 0.007 | 23 | 29 | 1.36 | 0.14 |
| 22 | 23 | 0.825 | 0.073 | 23 | 30 | 0.426 | 0.043 |
| 22 | 24 | 4.27 | 0.37 | 23 | 31 | 0.107 | 0.011 |
| 22 | 25 | 7.72 | 0.67 | 23 | 32 | 0.022 | 0.002 |
| 22 | 26 | 8.23 | 0.72 | | | | |
| 22 | 27 | 5.14 | 0.45 | 24 | 24 | 0.140 | 0.014 |
| 22 | 28 | 2.55 | 0.22 | 24 | 25 | 1.28 | 0.13 |
| 22 | 29 | 0.835 | 0.074 | 24 | 26 | 5.65 | 0.56 |
| 22 | 30 | 0.254 | 0.023 | 24 | 27 | 10.4 | 1.0 |
| 22 | 31 | 0.071 | 0.010 | 24 | 28 | 9.40 | 0.94 |
| | | | | 24 | 29 | 4.52 | 0.45 |
| 23 | 23 | 0.034 | 0.003 | 24 | 30 | 1.78 | 0.18 |
| 23 | 24 | 0.549 | 0.048 | 24 | 31 | 0.493 | 0.049 |
| 23 | 25 | 3.00 | 0.26 | 24 | 32 | 0.145 | 0.015 |
| 23 | 26 | 7.16 | 0.61 | 24 | 33 | 0.034 | 0.004 |
| 23 | 27 | 8.09 | 0.69 | 24 | 34 | 0.007 | 0.001 |
| 23 | 28 | 5.70 | 0.49 | | | | |
| 23 | 29 | 2.69 | 0.23 | 25 | 25 | 0.077 | 0.0079 |
| 23 | 30 | 1.21 | 0.10 | 25 | 26 | 0.895 | 0.089 |
| 23 | 31 | 0.379 | 0.033 | 25 | 27 | 4.59 | 0.46 |
| 23 | 32 | 0.128 | 0.012 | 25 | 28 | 9.73 | 0.97 |
| 24 | 24 | 0.031 | 0.003 | 25 | 29 | 9.24 | 0.92 |
| 24 | 25 | 0.423 | 0.036 | 25 | 30 | 5.51 | 0.55 |
| 24 | 26 | 2.64 | 0.22 | 25 | 31 | 1.93 | 0.19 |
| 24 | 27 | 6.46 | 0.54 | 25 | 32 | 0.711 | 0.071 |
| 24 | 28 | 8.17 | 0.69 | 25 | 33 | 0.185 | 0.019 |
| 24 | 29 | 5.70 | 0.48 | 25 | 34 | 0.057 | 0.006 |
| 24 | 30 | 3.37 | 0.29 | 25 | 35 | 0.012 | 0.001 |
| 24 | 31 | 1.33 | 0.11 | 25 | 36 | 0.0021 | 0.0003 |
| 24 | 32 | 0.565 | 0.048 | | | | |
| 24 | 33 | 0.190 | 0.017 | 26 | 25 | 0.008 | 0.002 |
| 24 | 34 | 0.058 | 0.008 | 26 | 26 | 0.059 | 0.009 |
| | | | | 26 | 27 | 0.661 | 0.103 |
| 25 | 25 | 0.017 | 0.002 | 26 | 28 | 3.55 | 0.55 |
| 25 | 26 | 0.278 | 0.024 | 26 | 29 | 7.99 | 1.25 |
| 25 | 27 | 1.97 | 0.16 | 26 | 30 | 9.53 | 1.49 |
| 25 | 28 | 5.67 | 0.47 | 26 | 31 | 5.61 | 0.88 |
| 25 | 29 | 7.45 | 0.62 | 26 | 32 | 2.67 | 0.42 |
| 25 | 30 | 6.44 | 0.53 | 26 | 33 | 0.813 | 0.127 |
| 25 | 31 | 3.50 | 0.29 | 26 | 34 | 0.269 | 0.042 |
| 25 | 32 | 1.93 | 0.16 | 26 | 35 | 0.064 | 0.010 |
| 25 | 33 | 0.735 | 0.062 | | | | |
| 25 | 34 | 0.287 | 0.024 | 27 | 27 | 0.031 | 0.005 |
| 25 | 35 | 0.091 | 0.009 | 27 | 28 | 0.417 | 0.060 |
| | | | | 27 | 29 | 2.53 | 0.37 |
| 26 | 26 | 0.013 | 0.002 | 27 | 30 | 7.31 | 1.05 |
| 26 | 27 | 0.199 | 0.017 | 27 | 31 | 9.15 | 1.32 |
| 26 | 28 | 1.55 | 0.13 | 27 | 32 | 7.38 | 1.07 |
| 26 | 29 | 4.80 | 0.39 | 27 | 33 | 3.22 | 0.47 |
| 26 | 30 | 8.01 | 0.65 | 27 | 34 | 1.23 | 0.18 |
| 26 | 31 | 6.57 | 0.53 | 27 | 35 | 0.378 | 0.055 |
| 26 | 32 | 4.56 | 0.37 | 27 | 36 | 0.103 | 0.015 |
| 26 | 33 | 2.26 | 0.18 | | | | |
| 26 | 34 | 1.05 | 0.09 | 28 | 27 | 0.006 | 0.001 |
| 26 | 35 | 0.390 | 0.032 | 28 | 28 | 0.021 | 0.003 |
| 26 | 36 | 0.119 | 0.011 | 28 | 29 | 0.240 | 0.032 |
| 26 | 37 | 0.037 | 0.004 | 28 | 30 | 1.69 | 0.22 |
| | | | | 28 | 31 | 5.75 | 0.76 |
| 27 | 27 | 0.008 | 0.001 | 28 | 32 | 9.83 | 1.31 |
| 27 | 28 | 0.110 | 0.009 | 28 | 33 | 8.27 | 1.10 |
| 27 | 29 | 0.97 | 0.08 | 28 | 34 | 4.75 | 0.63 |
| 27 | 30 | 3.95 | 0.31 | 28 | 35 | 1.66 | 0.22 |
| 27 | 31 | 6.88 | 0.55 | 28 | 36 | 0.544 | 0.072 |
| 27 | 32 | 7.56 | 0.60 | 28 | 37 | --- | --- |
| 27 | 33 | 4.92 | 0.39 | 28 | 38 | 0.042 | 0.006 |
| 27 | 34 | 2.99 | 0.24 | | | | |
| 27 | 35 | 1.34 | 0.11 | 29 | 29 | 0.012 | 0.002 |
| 27 | 36 | 0.572 | 0.046 | 29 | 30 | 0.130 | 0.015 |
| 27 | 37 | 0.194 | 0.016 | 29 | 31 | 1.04 | 0.12 |
| 27 | 38 | 0.078 | 0.009 | 29 | 32 | 4.31 | 0.51 |
| | | | | 29 | 33 | 8.44 | 0.98 |
| 28 | 29 | 0.051 | 0.005 | 29 | 34 | 8.94 | 1.04 |
| 28 | 30 | 0.577 | 0.046 | 29 | 35 | 5.45 | 0.63 |
| 28 | 31 | 2.76 | 0.22 | 29 | 36 | 2.44 | 0.28 |
| 28 | 32 | 6.54 | 0.51 | 29 | 37 | 0.819 | 0.096 |
| 28 | 33 | 7.69 | 0.60 | 29 | 38 | 0.279 | 0.033 |
| 28 | 34 | 6.25 | 0.49 | 29 | 39 | 0.061 | 0.007 |
| 28 | 35 | 3.58 | 0.28 | | | | |
| 28 | 36 | 1.87 | 0.15 | 30 | 31 | 0.098 | 0.011 |
| 28 | 37 | 0.786 | 0.062 | 30 | 32 | 0.836 | 0.091 |
| 28 | 38 | 0.284 | 0.023 | 30 | 33 | 3.80 | 0.41 |
| 28 | 39 | 0.088 | 0.008 | 30 | 34 | 8.73 | 0.95 |
| | | | | 30 | 35 | 10.5 | 1.1 |
| 29 | 31 | 0.291 | 0.023 | 30 | 36 | 7.26 | 0.79 |
| 29 | 32 | 1.79 | 0.14 | 30 | 37 | 3.60 | 0.39 |
| 29 | 33 | 4.96 | 0.39 | 30 | 38 | 1.54 | 0.17 |
| 29 | 34 | 7.54 | 0.59 | 30 | 39 | --- | --- |
| 29 | 35 | 6.44 | 0.50 | 30 | 40 | 0.104 | 0.012 |
| 29 | 36 | 4.56 | 0.36 | 30 | 41 | 0.020 | 0.002 |
| 29 | 37 | 2.52 | 0.20 | 30 | 42 | 0.0034 | 0.0005 |
| 29 | 38 | 1.17 | 0.09 | | | | |
| 29 | 39 | 0.455 | 0.036 | 31 | 32 | 0.055 | 0.006 |
| 29 | 40 | 0.162 | 0.013 | 31 | 33 | 0.489 | 0.048 |
| 29 | 41 | 0.051 | 0.005 | 31 | 34 | 2.55 | 0.25 |
| | | | | 31 | 35 | 6.95 | 0.68 |
| 30 | 31 | 0.014 | 0.002 | 31 | 36 | 10.3 | 1.0 |
| 30 | 32 | 0.207 | 0.017 | 31 | 37 | 8.33 | 0.82 |
| 30 | 33 | 1.36 | 0.11 | 31 | 38 | 4.32 | 0.42 |
| 30 | 34 | 4.54 | 0.35 | 31 | 39 | 2.29 | 0.23 |
| 30 | 35 | 7.55 | 0.59 | 31 | 40 | 0.861 | 0.088 |
| 30 | 36 | 7.56 | 0.59 | 31 | 41 | 0.272 | 0.028 |
| 30 | 37 | 5.34 | 0.42 | 31 | 42 | 0.041 | 0.004 |
| 30 | 38 | 3.19 | 0.25 | 32 | 33 | 0.044 | 0.004 |
| 30 | 39 | 1.59 | 0.12 | 32 | 34 | 0.371 | 0.036 |
| 30 | 40 | 0.681 | 0.054 | 32 | 35 | 2.12 | 0.20 |
| 30 | 41 | 0.248 | 0.020 | 32 | 36 | 6.61 | 0.63 |
| 30 | 42 | 0.084 | 0.008 | 32 | 37 | 11.1 | 1.1 |
| 30 | 43 | 0.018 | 0.003 | 32 | 38 | 10.7 | 1.0 |
| | | | | 32 | 39 | 5.86 | 0.56 |
| 31 | 33 | 0.107 | 0.008 | 32 | 40 | 2.99 | 0.29 |
| 31 | 34 | 0.782 | 0.058 | 32 | 41 | 1.06 | 0.10 |
| 31 | 35 | 3.14 | 0.23 | 32 | 42 | 0.331 | 0.034 |
| 31 | 36 | 6.43 | 0.47 | 32 | 43 | 0.084 | 0.008 |
| 31 | 37 | 7.61 | 0.56 | | | | |
| 31 | 38 | 6.13 | 0.45 | 33 | 34 | 0.035 | 0.003 |
| 31 | 39 | 4.06 | 0.30 | 33 | 35 | 0.240 | 0.017 |
| 31 | 40 | 2.27 | 0.17 | 33 | 36 | 1.41 | 0.10 |
| 31 | 41 | 0.974 | 0.072 | 33 | 37 | 5.08 | 0.36 |
| 31 | 42 | 0.422 | 0.031 | 33 | 38 | 10.4 | 0.7 |
| 31 | 43 | 0.146 | 0.012 | 33 | 39 | 11.1 | 0.8 |
| | | | | 33 | 40 | 7.54 | 0.54 |
| 32 | 34 | 0.068 | 0.006 | 33 | 41 | 3.49 | 0.25 |
| 32 | 35 | 0.552 | 0.041 | 33 | 42 | 1.54 | 0.11 |
| 32 | 36 | 2.60 | 0.19 | 33 | 43 | 0.389 | 0.031 |
| 32 | 37 | 6.12 | 0.44 | 33 | 44 | 0.159 | 0.013 |
| 32 | 38 | 8.26 | 0.60 | 33 | 45 | 0.018 | 0.002 |
| 32 | 39 | 7.04 | 0.51 | 33 | 46 | 0.0035 | 0.0004 |
| 32 | 40 | 5.02 | 0.36 | 34 | 35 | 0.022 | 0.002 |
| 32 | 41 | 2.85 | 0.21 | 34 | 36 | 0.184 | 0.013 |
| 32 | 42 | 1.41 | 0.10 | 34 | 37 | 1.18 | 0.08 |
| 32 | 43 | 0.58 | 0.04 | 34 | 38 | 4.81 | 0.34 |
| 32 | 44 | 0.211 | 0.016 | 34 | 39 | 10.3 | 0.7 |
| | | | | 34 | 40 | 13.6 | 0.9 |
| 33 | 35 | 0.0341 | 0.0030 | 34 | 41 | 9.33 | 0.66 |
| 33 | 36 | 0.342 | 0.025 | 34 | 42 | 5.04 | 0.36 |
| 33 | 37 | 1.72 | 0.12 | 34 | 43 | 1.98 | 0.15 |
| 33 | 38 | 5.10 | 0.37 | 34 | 44 | --- | --- |
| 33 | 39 | 7.79 | 0.56 | 34 | 45 | 0.231 | 0.019 |
| 33 | 40 | 8.15 | 0.59 | 34 | 46 | --- | --- |
| 33 | 41 | 5.79 | 0.42 | 34 | 47 | 0.0056 | 0.0005 |
| 33 | 42 | 3.83 | 0.28 | | | | |
| 33 | 43 | 1.89 | 0.14 | 35 | 36 | 0.013 | 0.001 |
| 33 | 44 | 0.914 | 0.066 | 35 | 37 | 0.119 | 0.009 |
| 33 | 45 | 0.318 | 0.023 | 35 | 38 | 0.798 | 0.057 |
| 33 | 46 | 0.140 | 0.016 | 35 | 39 | 3.46 | 0.25 |
| 33 | 47 | 0.041 | 0.004 | 35 | 40 | 9.27 | 0.66 |
| | | | | 35 | 41 | 13.0 | 0.9 |
| 34 | 36 | 0.023 | 0.002 | 35 | 42 | 11.6 | 0.8 |
| 34 | 37 | 0.244 | 0.018 | 35 | 43 | 5.53 | 0.39 |
| 34 | 38 | 1.44 | 0.10 | 35 | 44 | 2.97 | 0.21 |
| 34 | 39 | 4.32 | 0.31 | 35 | 45 | 0.973 | 0.071 |
| 34 | 40 | 8.16 | 0.58 | 35 | 46 | 0.274 | 0.022 |
| 34 | 41 | 8.45 | 0.60 | 35 | 47 | 0.133 | 0.013 |
| 34 | 42 | 7.08 | 0.50 | 35 | 48 | 0.009 | 0.001 |
| 34 | 43 | 4.44 | 0.32 | 36 | 38 | 0.079 | 0.006 |
| 34 | 44 | 2.63 | 0.19 | 36 | 39 | 0.639 | 0.045 |
| 34 | 45 | 1.22 | 0.09 | 36 | 40 | 3.09 | 0.22 |
| 34 | 46 | 0.513 | 0.037 | 36 | 41 | 8.73 | 0.62 |
| 34 | 47 | 0.195 | 0.015 | 36 | 42 | 14.6 | 1.0 |
| 34 | 48 | 0.056 | 0.005 | 36 | 42 | 14.6 | 1.0 |
| | | | | 36 | 43 | 13.2 | 0.9 |
| 35 | 37 | 0.012 | 0.001 | 36 | 44 | 8.53 | 0.60 |
| 35 | 38 | 0.126 | 0.009 | 36 | 45 | 3.72 | 0.27 |
| 35 | 39 | 0.873 | 0.063 | 36 | 46 | 1.24 | 0.09 |
| 35 | 40 | 3.39 | 0.24 | 36 | 47 | 0.329 | 0.025 |
| 35 | 41 | 6.97 | 0.49 | 36 | 48 | 0.080 | 0.006 |
| 35 | 42 | 9.10 | 0.65 | | | | |
| 35 | 43 | 7.46 | 0.53 | 37 | 39 | 0.047 | 0.004 |
| 35 | 44 | 5.88 | 0.42 | 37 | 40 | 0.393 | 0.028 |
| 35 | 45 | 3.33 | 0.24 | 37 | 41 | 2.18 | 0.15 |
| 35 | 46 | 1.78 | 0.13 | 37 | 42 | 7.33 | 0.52 |
| 35 | 47 | 0.766 | 0.055 | 37 | 43 | 13.4 | 0.9 |
| 35 | 48 | 0.285 | 0.021 | 37 | 44 | 15.5 | 1.1 |
| 35 | 49 | 0.102 | 0.009 | 37 | 45 | 10.2 | 0.7 |
| | | | | 37 | 46 | 5.06 | 0.36 |
| 36 | 39 | 0.091 | 0.007 | 37 | 47 | 1.72 | 0.12 |
| 36 | 40 | 0.686 | 0.049 | 37 | 48 | 0.380 | 0.030 |
| 36 | 41 | 2.79 | 0.20 | 37 | 49 | 0.126 | 0.009 |
| 36 | 42 | 6.69 | 0.47 | 37 | 50 | 0.017 | 0.002 |
| 36 | 43 | 8.89 | 0.63 | | | | |
| 36 | 44 | 9.09 | 0.64 | 38 | 40 | 0.034 | 0.003 |
| 36 | 45 | 6.68 | 0.47 | 38 | 41 | 0.296 | 0.021 |
| 36 | 46 | 4.36 | 0.31 | 38 | 42 | 1.74 | 0.12 |
| 36 | 47 | 2.38 | 0.17 | 38 | 43 | 6.37 | 0.45 |
| 36 | 48 | 1.09 | 0.08 | 38 | 44 | 13.9 | 0.9 |
| 36 | 49 | 0.419 | 0.030 | 38 | 45 | 17.2 | 1.2 |
| 36 | 50 | 0.136 | 0.010 | 38 | 46 | 13.7 | 1.0 |
| 36 | 51 | 0.043 | 0.007 | 38 | 47 | 6.69 | 0.47 |
| | | | | 38 | 48 | 2.70 | 0.19 |
| 37 | 40 | 0.044 | 0.005 | 38 | 49 | --- | --- |
| 37 | 41 | 0.391 | 0.028 | 38 | 50 | 0.194 | 0.014 |
| 37 | 42 | 1.95 | 0.14 | 38 | 51 | 0.060 | 0.007 |
| 37 | 43 | 5.29 | 0.38 | | | | |
| 37 | 44 | 9.05 | 0.64 | 39 | 42 | 0.178 | 0.013 |
| 37 | 45 | 9.28 | 0.66 | 39 | 43 | 1.20 | 0.09 |
| 37 | 46 | 8.17 | 0.58 | 39 | 44 | 4.93 | 0.35 |
| 37 | 47 | 5.43 | 0.38 | 39 | 45 | 12.3 | 0.9 |
| 37 | 48 | 3.18 | 0.23 | 39 | 46 | 18.2 | 1.3 |
| 37 | 49 | 1.48 | 0.11 | 39 | 47 | 16.2 | 1.1 |
| 37 | 50 | 0.595 | 0.043 | 39 | 48 | 9.47 | 0.67 |
| 37 | 51 | 0.222 | 0.016 | 39 | 49 | 3.52 | 0.25 |
| 37 | 52 | 0.057 | 0.005 | 39 | 50 | 1.13 | 0.08 |
| | | | | 39 | 51 | 0.246 | 0.019 |
| 38 | 41 | 0.038 | 0.004 | 39 | 52 | 0.064 | 0.005 |
| 38 | 42 | 0.266 | 0.019 | | | | |
| 38 | 43 | 1.47 | 0.11 | 40 | 43 | 0.132 | 0.009 |
| 38 | 44 | 4.67 | 0.33 | 40 | 44 | 0.899 | 0.064 |
| 38 | 45 | 8.69 | 0.62 | 40 | 45 | 4.13 | 0.29 |
| 38 | 46 | 10.4 | 0.7 | 40 | 46 | 11.5 | 0.8 |
| 38 | 47 | 9.51 | 0.67 | 40 | 47 | 19.4 | 1.4 |
| 38 | 48 | 6.86 | 0.49 | 40 | 48 | 19.6 | 1.4 |
| 38 | 49 | 4.00 | 0.28 | 40 | 49 | 12.0 | 0.9 |
| 38 | 50 | 1.92 | 0.14 | 40 | 50 | 4.75 | 0.34 |
| 38 | 51 | 0.867 | 0.062 | 40 | 51 | 1.27 | 0.09 |
| 38 | 52 | 0.340 | 0.025 | 40 | 52 | 0.385 | 0.033 |
| 38 | 53 | 0.140 | 0.011 | 40 | 53 | 0.127 | 0.009 |
| 38 | 54 | 0.052 | 0.006 | | | | |
| | | | | 41 | 44 | 0.078 | 0.006 |
| 39 | 43 | 0.159 | 0.012 | 41 | 45 | 0.615 | 0.044 |
| 39 | 44 | 0.973 | 0.069 | 41 | 46 | 3.16 | 0.22 |
| 39 | 45 | 3.62 | 0.26 | 41 | 47 | 10.0 | 0.7 |
| 39 | 46 | 8.04 | 0.57 | 41 | 48 | 19.3 | 1.4 |
| 39 | 47 | 10.9 | 0.8 | 41 | 49 | 21.5 | 1.5 |
| 39 | 48 | 10.9 | 0.8 | 41 | 50 | 14.2 | 1.0 |
| 39 | 49 | 8.09 | 0.57 | 41 | 51 | 5.78 | 0.41 |
| 39 | 50 | 4.89 | 0.35 | 41 | 52 | 1.99 | 0.14 |
| 39 | 51 | 2.50 | 0.18 | 41 | 53 | --- | --- |
| 39 | 52 | 1.21 | 0.09 | 41 | 54 | 0.217 | 0.016 |
| 39 | 53 | 0.551 | 0.039 | 41 | 55 | 0.045 | 0.004 |
| 39 | 54 | 0.245 | 0.018 | | | | |
| 39 | 55 | 0.085 | 0.007 | 42 | 46 | 0.438 | 0.031 |
| | | | | 42 | 47 | 2.53 | 0.18 |
| 40 | 44 | 0.088 | 0.007 | 42 | 48 | 9.13 | 0.65 |
| 40 | 45 | 0.690 | 0.049 | 42 | 49 | 19.3 | 1.4 |
| 40 | 46 | 2.93 | 0.21 | 42 | 50 | 23.2 | 1.6 |
| 40 | 47 | 7.31 | 0.52 | 42 | 51 | 14.9 | 1.1 |
| 40 | 48 | 11.4 | 0.8 | 42 | 52 | 7.61 | 0.54 |
| 40 | 49 | 11.9 | 0.8 | 42 | 53 | 2.71 | 0.19 |
| 40 | 50 | 9.39 | 0.66 | 42 | 54 | --- | --- |
| 40 | 51 | 5.72 | 0.41 | 42 | 55 | 0.346 | 0.026 |
| 40 | 52 | 3.34 | 0.24 | 42 | 56 | 0.093 | 0.007 |
| 40 | 53 | 1.78 | 0.13 | 42 | 57 | 0.017 | 0.001 |
| 40 | 54 | 0.874 | 0.062 | | | | |
| 40 | 55 | 0.411 | 0.030 | 43 | 46 | 0.035 | 0.003 |
| 40 | 56 | 0.172 | 0.013 | 43 | 47 | 0.308 | 0.022 |
| 40 | 57 | 0.078 | 0.007 | 43 | 48 | 1.94 | 0.14 |
| | | | | 43 | 49 | 8.05 | 0.57 |
| 41 | 45 | 0.055 | 0.005 | 43 | 50 | 18.7 | 1.3 |
| 41 | 46 | 0.433 | 0.032 | 43 | 51 | 22.3 | 1.6 |
| 41 | 47 | 2.18 | 0.16 | 43 | 52 | 17.6 | 1.2 |
| 41 | 48 | 6.47 | 0.46 | 43 | 53 | 9.78 | 0.69 |
| 41 | 49 | 11.0 | 0.8 | 43 | 54 | 4.60 | 0.33 |
| 41 | 50 | 12.2 | 0.9 | 43 | 55 | 1.39 | 0.10 |
| 41 | 51 | 9.69 | 0.69 | 43 | 57 | 0.186 | 0.014 |
| 41 | 52 | 6.96 | 0.49 | 43 | 58 | 0.042 | 0.003 |
| 41 | 53 | 4.44 | 0.31 | 43 | 59 | 0.008 | 0.001 |
| 41 | 54 | 2.71 | 0.19 | | | | |
| 41 | 55 | 1.44 | 0.10 | 44 | 48 | 0.221 | 0.016 |
| 41 | 56 | 0.724 | 0.052 | 44 | 49 | 1.56 | 0.11 |
| 41 | 57 | 0.329 | 0.024 | 44 | 50 | 7.06 | 0.50 |
| 41 | 58 | 0.148 | 0.012 | 44 | 51 | 16.5 | 1.2 |
| 1 | 59 | 0.046 | 0.006 | 44 | 52 | 23.6 | 1.7 |
| | | | | 44 | 53 | 19.8 | 1.4 |
| 42 | 46 | 0.039 | 0.005 | 44 | 54 | 13.3 | 0.9 |
| 42 | 47 | 0.298 | 0.022 | 44 | 55 | 6.58 | 0.5 |
| 42 | 48 | 1.66 | 0.12 | 44 | 56 | 1.67 | 0.13 |
| 42 | 49 | 5.54 | 0.39 | 44 | 57 | --- | --- |
| 42 | 50 | 10.4 | 0.7 | 44 | 58 | 0.340 | 0.025 |
| 42 | 51 | 11.4 | 0.8 | 44 | 59 | 0.087 | 0.007 |
| 42 | 52 | 10.6 | 0.8 | 44 | 60 | 0.013 | 0.001 |
| 42 | 53 | 8.24 | 0.58 | | | | |
| 42 | 54 | 5.97 | 0.42 | 45 | 48 | 0.015 | 0.001 |
| 42 | 55 | 3.82 | 0.27 | 45 | 49 | 0.142 | 0.010 |
| 42 | 56 | 2.22 | 0.16 | 45 | 50 | 1.16 | 0.08 |
| 42 | 57 | 1.16 | 0.08 | 45 | 51 | 5.42 | 0.38 |
| 42 | 58 | 0.585 | 0.042 | 45 | 52 | 14.7 | 1.0 |
| 42 | 59 | 0.257 | 0.019 | 45 | 53 | 22.6 | 1.6 |
| 42 | 60 | 0.100 | 0.010 | 45 | 54 | 23.4 | 1.7 |
| | | | | 45 | 55 | 16.3 | 1.2 |
| 43 | 48 | 0.186 | 0.014 | 45 | 56 | 9.60 | 0.68 |
| 43 | 49 | 1.27 | 0.09 | 45 | 57 | 4.62 | 0.34 |
| 43 | 50 | 4.51 | 0.32 | 45 | 58 | --- | --- |
| 43 | 51 | 8.64 | 0.61 | 45 | 59 | 0.582 | 0.042 |
| 43 | 52 | 10.9 | 0.8 | 45 | 60 | 0.169 | 0.012 |
| 43 | 53 | 10.8 | 0.77 | 45 | 61 | 0.028 | 0.002 |
| 43 | 54 | 9.99 | 0.71 | 45 | 62 | 0.0077 | 0.0007 |
| 43 | 55 | 7.57 | 0.54 | 46 | 50 | 0.102 | 0.007 |
| 43 | 56 | 5.32 | 0.38 | 46 | 51 | 0.804 | 0.057 |
| 43 | 57 | 3.24 | 0.23 | 46 | 52 | 4.11 | 0.29 |
| 43 | 58 | 1.94 | 0.14 | 46 | 53 | 12.2 | 0.9 |
| 43 | 59 | 0.999 | 0.071 | 46 | 54 | 23.1 | 1.6 |
| 43 | 60 | 0.538 | 0.039 | 46 | 55 | 25.8 | 1.8 |
| 43 | 61 | 0.211 | 0.016 | 46 | 56 | 21.6 | 1.5 |
| 43 | 62 | 0.099 | 0.010 | 46 | 57 | 13.1 | 0.9 |
| | | | | 46 | 58 | 6.20 | 0.44 |
| 44 | 49 | 0.129 | 0.010 | 46 | 59 | 1.97 | 0.14 |
| 44 | 50 | 0.946 | 0.068 | 46 | 60 | 0.805 | 0.087 |
| 44 | 51 | 3.26 | 0.23 | 46 | 61 | 0.315 | 0.023 |
| 44 | 52 | 7.29 | 0.52 | 46 | 62 | 0.086 | 0.007 |
| 44 | 53 | 10.0 | 0.7 | 46 | 63 | 0.011 | 0.001 |
| 44 | 54 | 11.7 | 0.8 | | | | |
| 44 | 55 | 11.4 | 0.8 | 47 | 51 | 0.055 | 0.004 |
| 44 | 56 | 9.46 | 0.67 | 47 | 52 | 0.499 | 0.036 |
| 44 | 57 | 7.05 | 0.50 | 47 | 53 | 2.69 | 0.19 |
| 44 | 58 | 4.72 | 0.33 | 47 | 54 | 9.70 | 0.69 |
| 44 | 59 | 2.82 | 0.20 | 47 | 55 | 20.1 | 1.4 |
| 44 | 60 | 1.63 | 0.12 | 47 | 56 | 28.2 | 2.0 |
| 44 | 61 | 0.852 | 0.061 | 47 | 57 | 25.7 | 1.8 |
| 44 | 62 | 0.418 | 0.031 | 47 | 58 | 17.8 | 1.3 |
| 44 | 63 | 0.187 | 0.015 | 47 | 59 | 9.20 | 0.65 |
| 44 | 64 | 0.077 | 0.008 | 47 | 60 | 3.92 | 0.28 |
| | | | | 47 | 61 | --- | --- |
| 45 | 50 | 0.077 | 0.006 | 47 | 62 | 0.566 | 0.040 |
| 45 | 51 | 0.584 | 0.042 | 47 | 63 | 0.169 | 0.013 |
| 45 | 52 | 2.33 | 0.17 | 47 | 64 | 0.036 | 0.003 |
| 45 | 53 | 5.65 | 0.40 | 47 | 65 | 0.0072 | 0.0006 |
| 45 | 54 | 9.24 | 0.65 | | | | |
| 45 | 55 | 11.4 | 0.8 | 48 | 52 | 0.028 | 0.002 |
| 45 | 56 | 12.6 | 0.9 | 48 | 53 | 0.278 | 0.020 |
| 45 | 57 | 11.2 | 0.8 | 48 | 54 | 1.71 | 0.12 |
| 45 | 58 | 9.17 | 0.65 | 48 | 55 | 6.88 | 0.49 |
| 45 | 59 | 6.57 | 0.47 | 48 | 56 | 17.7 | 1.3 |
| 45 | 60 | 4.47 | 0.32 | 48 | 57 | 28.7 | 2.0 |
| 45 | 61 | 2.63 | 0.19 | 48 | 58 | 31.5 | 2.2 |
| 45 | 62 | 1.56 | 0.11 | 48 | 59 | 22.9 | 1.63 |
| 45 | 63 | 0.789 | 0.057 | 48 | 60 | 14.1 | 1.0 |
| 45 | 64 | 0.425 | 0.031 | 48 | 61 | 6.34 | 0.45 |
| 45 | 65 | 0.171 | 0.014 | 48 | 62 | --- | --- |
| 45 | 66 | 0.070 | 0.008 | 48 | 63 | 1.09 | 0.08 |
| | | | | 48 | 64 | 0.274 | 0.020 |
| 46 | 51 | 0.024 | 0.003 | 48 | 66 | 0.018 | 0.002 |
| 46 | 52 | 0.303 | 0.022 | 48 | 67 | 0.0032 | 0.0005 |
| 46 | 53 | 1.45 | 0.10 | | | | |
| 46 | 55 | 7.78 | 0.55 | 49 | 53 | 0.0124 | 0.0011 |
| 46 | 56 | 11.1 | 0.8 | 49 | 54 | 0.124 | 0.009 |
| 46 | 57 | 12.9 | 0.9 | 49 | 55 | 0.861 | 0.061 |
| 46 | 58 | 12.7 | 0.9 | 49 | 56 | 4.20 | 0.30 |
| 46 | 59 | 11.1 | 0.8 | 49 | 57 | 12.8 | 0.9 |
| 46 | 60 | 8.73 | 0.62 | 49 | 58 | 26.5 | 1.9 |
| 46 | 61 | 6.20 | 0.44 | 49 | 59 | 33.5 | 2.4 |
| 46 | 62 | 4.11 | 0.29 | 49 | 60 | 29.4 | 2.1 |
| 46 | 63 | 2.51 | 0.18 | 49 | 61 | 19.7 | 1.4 |
| 46 | 64 | 1.46 | 0.10 | 49 | 62 | 9.55 | 0.68 |
| 46 | 65 | 0.742 | 0.054 | 49 | 63 | --- | --- |
| 46 | 66 | 0.402 | 0.030 | 49 | 64 | 1.54 | 0.11 |
| 46 | 67 | 0.144 | 0.013 | 49 | 65 | 0.729 | 0.052 |
| 46 | 68 | 0.072 | 0.008 | 49 | 66 | 0.218 | 0.016 |
| | | | | 49 | 67 | 0.033 | 0.003 |
| 47 | 53 | 0.159 | 0.012 | 49 | 68 | 0.0094 | 0.0008 |
| 47 | 54 | 0.859 | 0.062 | | | | |
| 47 | 55 | 2.79 | 0.20 | 50 | 55 | 0.048 | 0.004 |
| 47 | 56 | 6.11 | 0.43 | 50 | 56 | 0.420 | 0.030 |
| 47 | 57 | 9.49 | 0.67 | 50 | 57 | 2.38 | 0.17 |
| 47 | 58 | 12.6 | 0.9 | 50 | 58 | 8.64 | 0.61 |
| 47 | 59 | 13.4 | 0.9 | 50 | 59 | 22.6 | 1.6 |
| 47 | 60 | 13.1 | 0.9 | 50 | 60 | 35.9 | 2.5 |
| 47 | 61 | 11.0 | 0.8 | 50 | 61 | 37.7 | 2.7 |
| 47 | 62 | 8.81 | 0.62 | 50 | 62 | 31.0 | 2.2 |
| 47 | 63 | 6.17 | 0.44 | 50 | 63 | 22.4 | 1.6 |
| 47 | 64 | 4.28 | 0.30 | 50 | 64 | 6.51 | 0.46 |
| 47 | 65 | 2.58 | 0.18 | 50 | 65 | --- | --- |
| 47 | 66 | 1.59 | 0.11 | 50 | 66 | 1.66 | 0.12 |
| 47 | 67 | 0.872 | 0.063 | 50 | 67 | 0.581 | 0.042 |
| 47 | 68 | 0.458 | 0.034 | 50 | 68 | 0.108 | 0.008 |
| 47 | 69 | 0.224 | 0.018 | 50 | 69 | 0.022 | 0.002 |
| | | | | | | | |
| 48 | 54 | 0.067 | 0.006 | 51 | 56 | 0.014 | 0.001 |
| 48 | 55 | 0.408 | 0.030 | 51 | 57 | 0.144 | 0.010 |
| 48 | 56 | 1.73 | 0.12 | 51 | 58 | 0.927 | 0.066 |
| 48 | 57 | 4.43 | 0.31 | 51 | 59 | 3.95 | 0.28 |
| 48 | 58 | 8.09 | 0.57 | 51 | 60 | 13.3 | 0.9 |
| 48 | 59 | 11.6 | 0.8 | 51 | 61 | 26.7 | 1.9 |
| 48 | 60 | 13.9 | 1.0 | 51 | 62 | 36.9 | 2.6 |
| 48 | 61 | 14.5 | 1.0 | 51 | 63 | 36.1 | 2.6 |
| 48 | 62 | 13.5 | 1.0 | 51 | 64 | 28.6 | 2.0 |
| 48 | 63 | 11.5 | 0.9 | 51 | 65 | --- | --- |
| 48 | 64 | 8.99 | 0.64 | 51 | 66 | --- | --- |
| 48 | 65 | 6.51 | 0.46 | 51 | 67 | 3.88 | 0.27 |
| 48 | 66 | 4.58 | 0.32 | 51 | 68 | 1.73 | 0.12 |
| 48 | 67 | 2.90 | 0.21 | 51 | 69 | 0.417 | 0.030 |
| 48 | 68 | 1.79 | 0.13 | 51 | 70 | 0.087 | 0.006 |
| 48 | 69 | 1.03 | 0.07 | | | | |
| 48 | 70 | 0.576 | 0.043 | 52 | 58 | 0.059 | 0.005 |
| 48 | 71 | 0.263 | 0.022 | 52 | 59 | 0.438 | 0.032 |
| 48 | 72 | --- | --- | 52 | 60 | 2.26 | 0.16 |
| 48 | 73 | --- | --- | 52 | 61 | 8.74 | 0.62 |
| 48 | 74 | 0.033 | 0.004 | 52 | 62 | 22.5 | 1.6 |
| 48 | 75 | - | - | 52 | 63 | 36.7 | 2.6 |
| 48 | 76 | 0.0044 | 0.0004 | 52 | 64 | 45.2 | 3.2 |
| 48 | 77 | 0.0013 | 0.0002 | 52 | 65 | 41.6 | 2.9 |
| | | | | 52 | 66 | 24.2 | 1.7 |
| 49 | 56 | 0.161 | 0.012 | 52 | 67 | --- | --- |
| 49 | 57 | 0.834 | 0.060 | 52 | 68 | 12.4 | 0.9 |
| 49 | 58 | 2.73 | 0.19 | 52 | 69 | 6.08 | 0.43 |
| 49 | 59 | 5.87 | 0.42 | 52 | 70 | 2.23 | 0.16 |
| 49 | 60 | 9.75 | 0.69 | | | | |
| 49 | 61 | 12.9 | 0.9 | 53 | 59 | 0.018 | 0.002 |
| 49 | 62 | 15.1 | 1.1 | 53 | 60 | 0.145 | 0.011 |
| 49 | 63 | 15.2 | 1.1 | 53 | 61 | 0.973 | 0.070 |
| 49 | 64 | 14.7 | 1.0 | 53 | 62 | 4.41 | 0.31 |
| 49 | 65 | 12.3 | 0.9 | 53 | 63 | 13.8 | 1.0 |
| 49 | 66 | 10.2 | 0.7 | 53 | 64 | 31.0 | 2.2 |
| 49 | 67 | 7.50 | 0.53 | 53 | 65 | 45.3 | 3.2 |
| 49 | 68 | 5.57 | 0.39 | 53 | 66 | 59.0 | 4.2 |
| 49 | 69 | 3.70 | 0.26 | 53 | 67 | 40.7 | 2.9 |
| 49 | 70 | 2.44 | 0.17 | 53 | 68 | --- | --- |
| 49 | 71 | 1.43 | 0.10 | 53 | 69 | --- | --- |
| 49 | 72 | 0.993 | 0.080 | 53 | 70 | 112 | 8 |
| 49 | 73 | 0.505 | 0.051 | | | | |
| 49 | 74 | --- | --- | 54 | 61 | 0.057 | 0.009 |
| 49 | 75 | 0.101 | 0.017 | 54 | 62 | 0.444 | 0.067 |
| 49 | 76 | 0.068 | 0.006 | 54 | 63 | 2.21 | 0.33 |
| 49 | 77 | 0.012 | 0.001 | 54 | 64 | 8.89 | 1.34 |
| 49 | 78 | 0.0057 | 0.0005 | 54 | 65 | 24.6 | 3.7 |
| | | | | 54 | 66 | 53.5 | 8.0 |
| | | | | 54 | 67 | 96.7 | 14.5 |
| 50 | 57 | 0.063 | 0.007 | 54 | 68 | 333 | 50 |
| 50 | 58 | 0.359 | 0.033 | 54 | 69 | 24.9 | 3.7 |
| 50 | 59 | 1.46 | 0.13 | | | | |
| 50 | 60 | 3.84 | 0.35 | 55 | 63 | 0.076 | 0.013 |
| 50 | 61 | 7.44 | 0.68 | 55 | 64 | 0.360 | 0.062 |
| 50 | 62 | 11.5 | 1.0 | 55 | 65 | 1.34 | 0.23 |
| 50 | 63 | 14.5 | 1.3 | 55 | 66 | 3.52 | 0.60 |
| 50 | 64 | 16.5 | 1.5 | 55 | 67 | 4.22 | 0.72 |
| 50 | 65 | 17.2 | 1.6 | 55 | 68 | 3.75 | 0.64 |
| 50 | 66 | 16.2 | 1.5 | 55 | 69 | 1.70 | 0.29 |
| 50 | 67 | 14.6 | 1.3 | | | | |
| 50 | 68 | 11.9 | 1.1 | | | | |
| 50 | 69 | 9.58 | 0.87 | | | | |
| 50 | 70 | 6.96 | 0.63 | | | | |
| 50 | 71 | 5.07 | 0.46 | | | | |
| 50 | 72 | 3.54 | 0.32 | | | | |
| 50 | 73 | 2.43 | 0.22 | | | | |
| 50 | 74 | 1.73 | 0.16 | | | | |
| 50 | 75 | --- | --- | | | | |
| 50 | 76 | --- | --- | | | | |
| 50 | 77 | 0.218 | 0.021 | | | | |
| 50 | 78 | --- | --- | | | | |
| 50 | 79 | 0.025 | 0.002 | | | | |
| 50 | 80 | 0.0064 | 0.0006 | | | | |
| | | | | | | | |
| 51 | 59 | 0.109 | 0.011 | | | | |
| 51 | 60 | 0.516 | 0.048 | | | | |
| 51 | 61 | 1.87 | 0.17 | | | | |
| 51 | 62 | 4.37 | 0.40 | | | | |

| | | | |
|---|---|---|---|
| 51 | 63 | 8.10 | 0.73 |
| 51 | 64 | 11.9 | 1.1 |
| 51 | 65 | 15.1 | 1.4 |
| 51 | 66 | 17.9 | 1.6 |
| 51 | 67 | 18.8 | 1.7 |
| 51 | 68 | 19.1 | 1.7 |
| 51 | 69 | 17.4 | 1.6 |
| 51 | 70 | 16.0 | 1.5 |
| 51 | 71 | 13.2 | 1.2 |
| 51 | 72 | 11.0 | 1.0 |
| 51 | 73 | 8.46 | 0.77 |
| 51 | 74 | 6.62 | 0.60 |
| 51 | 75 | 4.84 | 0.44 |
| 51 | 76 | 4.28 | 0.41 |
| 51 | 77 | 1.66 | 0.17 |
| 51 | 78 | 0.844 | 0.078 |
| 51 | 79 | 0.502 | 0.049 |
| 51 | 80 | 0.174 | 0.018 |
| 51 | 81 | 0.047 | 0.004 |
| 51 | 82 | 0.0070 | 0.0007 |
| | | | |
| 52 | 60 | 0.039 | 0.006 |
| 52 | 61 | 0.208 | 0.024 |
| 52 | 62 | 0.849 | 0.095 |
| 52 | 63 | 2.38 | 0.26 |
| 52 | 64 | 5.18 | 0.57 |
| 52 | 65 | 8.66 | 0.96 |
| 52 | 66 | 13.1 | 1.5 |
| 52 | 67 | 16.3 | 1.8 |
| 52 | 68 | 19.5 | 2.2 |
| 52 | 69 | 21.9 | 2.4 |
| 52 | 70 | 22.9 | 2.5 |
| 52 | 71 | 23.2 | 2.6 |
| 52 | 72 | 22.0 | 2.4 |
| 52 | 73 | 20.5 | 2.3 |
| 52 | 74 | 18.5 | 2.0 |
| 52 | 75 | 16.1 | 1.8 |
| 52 | 76 | 14.1 | 1.6 |
| 52 | 77 | 10.9 | 1.2 |
| 52 | 78 | 9.86 | 1.11 |
| 52 | 79 | ---- | --- |
| 52 | 80 | 3.56 | 0.39 |
| 52 | 81 | 0.525 | 0.059 |
| 52 | 82 | 0.319 | 0.035 |
| 52 | 83 | 0.0038 | 0.0005 |
| | | | |
| 53 | 62 | 0.032 | 0.005 |
| 53 | 63 | 0.277 | 0.037 |
| 53 | 64 | 0.966 | 0.127 |
| 53 | 65 | 2.39 | 0.31 |
| 53 | 66 | 5.07 | 0.66 |
| 53 | 67 | 7.92 | 1.03 |
| 53 | 68 | 11.9 | 1.6 |
| 53 | 69 | 15.6  | 2.0   |
| 53 | 70 | 20.5  | 2.7   |
| 53 | 71 | 23.1  | 3.0   |
| 53 | 72 | 28.7  | 3.7   |
| 53 | 73 | 28.5  | 3.7   |
| 53 | 74 | 35.2  | 4.6   |
| 53 | 75 | 32.2  | 4.2   |
| 53 | 76 | 38.3  | 5.0   |
| 53 | 77 | 33.9  | 4.4   |
| 53 | 78 | 44.5  | 5.8   |
| 53 | 79 | 37.4  | 4.9   |
| 53 | 80 | 31.7  | 4.3   |
| 53 | 81 | 26.6  | 3.5   |
| 53 | 82 | 30.6  | 3.9   |
| 54 | 64 | 0.075 | 0.012 |
| 54 | 65 | 0.294 | 0.045 |
| 54 | 66 | 0.864 | 0.130 |
| 54 | 67 | 1.83  | 0.28  |
| 54 | 68 | 3.66  | 0.55  |
| 54 | 69 | 5.63  | 0.85  |
| 54 | 70 | 9.07  | 1.36  |
| 54 | 71 | 11.7  | 1.8   |
| 54 | 72 | 18.0  | 2.7   |
| 54 | 73 | 20.1  | 3.0   |
| 54 | 74 | 27.5  | 4.1   |
| 54 | 75 | 32.5  | 4.9   |
| 54 | 76 | 44.0  | 6.6   |
| 54 | 77 | 57.0  | 8.6   |
| 54 | 78 | 95.8  | 14.4  |
| 54 | 79 | 207   | 31    |
| 54 | 80 | 656   | 98    |
| 54 | 81 | 2201  | 331   |
| 55 | 66 | 0.053 | 0.010 |
| 55 | 67 | 0.147 | 0.026 |
| 55 | 68 | 0.377 | 0.065 |
| 55 | 69 | 0.738 | 0.126 |
| 55 | 70 | 1.32  | 0.23  |
| 55 | 71 | 2.00  | 0.34  |
| 55 | 72 | 3.04  | 0.52  |
| 55 | 73 | 4.01  | 0.68  |
| 55 | 74 | 5.45  | 0.93  |
| 55 | 75 | 6.56  | 1.12  |
| 55 | 76 | 7.60  | 1.29  |
| 55 | 77 | 8.17  | 1.39  |
| 55 | 78 | 7.39  | 1.26  |
| 55 | 79 | 5.59  | 0.95  |
| 55 | 80 | 2.98  | 0.51  |
| 55 | 81 | 1.28  | 0.22  |
| 56 | 69 | 0.025 | 0.005 |
| 56 | 70 | 0.044 | 0.008 |

| 56 | 71 | 0.067 | 0.012 |
| 56 | 72 | 0.090 | 0.016 |
| 56 | 73 | 0.093 | 0.016 |
| 56 | 74 | 0.088 | 0.015 |
| 56 | 75 | 0.075 | 0.013 |
| 56 | 76 | 0.046 | 0.008 |
| 56 | 77 | 0.028 | |